\documentclass[journal]{IEEEtran}
\ifCLASSINFOpdf
\else
\fi
\usepackage{amssymb}
\usepackage{fancyhdr}
\usepackage{graphicx} 
\usepackage{epsfig}
\usepackage{multirow}
\usepackage[center]{caption}
\usepackage{subcaption}
\usepackage{setspace}
\usepackage{amsmath}
\usepackage{float}
\usepackage{subfig}
\usepackage{cite}
\usepackage{array}
\usepackage{verbatim}
\usepackage{graphicx} 
\usepackage{amsfonts}
\usepackage{MnSymbol}%
\twocolumn
\date{}
\begin{document}
\title{Exact Least Squares Algorithm for Signal Matched Multirate Whitening Filter
Bank:Part I}
\author{Binish~Fatimah and S.~D.~Joshi~
\thanks{The authors are with the Department
of Electrical  Engineering, Indian Institute of Technology, New Delhi 110016 India e-mail: (binish.fatimah@ee.iitd.ac.in)(sdjoshi@ee.iitd.ac.in).}}



\maketitle
\begin{abstract}
In this paper, we define a concept of signal matched multirate whitening filter bank
 which provides an optimum coding gain. This is achieved by whitening the outputs, of the analysis filter bank, within as well as across the channels, by solving a constrained prediction problem. We also present a fast time and order recursive least squares algorithm to obtain the vector output of the proposed analysis filter bank.
The recursive algorithm, developed here, gives rise to a lattice-like structure. Since the proposed signal matched analysis filter bank coefficients are not available directly, an order recursive algorithm is also  presented for estimating these from the lattice parameters. Simulation results are presented to validate the theory. It is also 
observed that the proposed algorithm can be used to whiten Gaussian/non-Gaussian processes with minimum as well as non-minimum phase.

\end{abstract}


\section{Introduction}
Whitening, a given sequence of random variables, is the most fundamental issue of concern in the compression process. The whitening process is intrinsically linked to the underlying model of the process. A number of approaches exist, in the literature, for whitening signals having different models \cite{Friedlander}. Whereas, the standard models such as Gaussian/non-Gaussian AR, MA, ARMA etc have been extensively  investigated in the literature, similar investigation, in the context of models based on multirate filter bank, still happens to be an area of current research activity. A number of researchers have proposed different ways for the design of multirate filter bank, addressing fundamentally the same issue of whitening a sequence.
 These filter banks are adapted to a given input signal or to its statistics. This concept has been investigated by many authors\cite{Nalbalwar_phd,ref1,Lu,weng,A.gupta,Delsarte}. In the applications involving pattern recognition, signal modeling, compression, sub-band coders, etc, they offer better results \cite{Nalbalwar_phd,ref3,ref4,weng}.

Delsarte et. al. \cite{Delsarte} designed a multiresolution transform,
adapted to a given stationary signal, such that the variance of detail signals is minimized at each resolution level. The parameters were optimized using a non-linear ``ring algorithm", which ensured their convergence only at a local minimum.
  A signal adapted M-channel biorthogonal filter bank of finite length is proposed by Lu et. al. in \cite{Lu}, such that it minimizes a coding gain related objective function, subject to perfect reconstruction condition. The non-linear constrained optimization problem was converted to a line search problem by parameterizing a first-order
approximation of the PR constraint. However, the proposed algorithm only satisfies perfect reconstruction condition approximately. Also the initial values, of the parameters, affect the performance of the algorithm. Tsatsanis et. al. \cite{ref1} designed a P-band orthogonal perfect reconstruction filter bank such that it minimizes the mean square error between the original signal and its low resolution version. And thus decomposes the input signal into its uncorrelated low-resolution principal components with decreasing variance. However, \cite{ref1} does not propose any FIR implementation and also the estimation of filter bank parameters is not discussed.

 Recently Weng and Vaidyanathan \cite{weng} proposed biorthogonal GTD (generalized triangular decomposition) filter banks for optimizing coding gain. The proposed filter bank is a cascade of an optimally orthonormal GTD SBC (Sub Band Coder) precoded with a set of filters depending on the input signal spectra. However they have restricted the filter bank with the condition of perfect reconstruction for every signal. Using the concept of Principal Component filter bank, Jhawar et al \cite{gadre} proposed an FIR PU (Para Unitary) filter bank of fixed length filters, for a uniformly decimated sub-band coder, maximizing the coding gain. However the results and the design presented were only for a fixed length filter bank and also the complexity of the algorithm increases many fold as the filter order increases.
 
 In this paper, we define a concept of signal matched whitening filter bank (SMWFB) which ensures de-correlation in time as well as across various channels and provides an optimum coding gain. It would be pertinent to mention here that a slightly different definition is proposed in \cite{Nalbalwar_phd}, which however does not guarantee de-correlation across channels. To obtain the required de-correlation both in time and across bands, we solve a constrained prediction problem. The constrained prediction problem is first proposed in a general context, in the form of a lemma, and is then used to define the required SMWFB, such that its outputs are constrained to be orthogonal in time and across channels. We define a Hilbert space framework to develop the geometrical counterpart of SMWFB, for the given data case. 
 Using the projection operator update relations, given in \cite{Friedlander} and \cite{joshiI}, we develop a time as well as order recursive least squares algorithm for the proposed SMWFB. Recursions of the algorithm  give rise to a lattice-ladder like structure. 
Although the algorithm developed here is for wide sense stationary and cyclo-stationary processes, its extension to the process with slowly time varying statistics is also straight forward. This can be achieved by simple exponential windowing technique conventionally used in the development of fast least squares lattice algorithms \cite{Friedlander}.

The least squares multirate whitening filter bank, discussed above, does not provide the filter bank coefficients directly but rather in terms of lattice parameters. So we develop a fast algorithm, for the computation of the filter bank coefficients in the least squares sense, by making use of pseudo-inverse update relation given in \cite{joshiII}.

This paper is organized as follows: In section II, we propose a lemma for a constrained projection problem.
Using this lemma we define the concept of signal matched multirate whitening filter bank, such that it gives optimum coding gain, in section III. In section IV, a Hilbert space framework is developed for the proposed problem which provides the given data interpretation of the Signal Matched Whitening Filter Bank, defined in the previous section. With the ``given data case" interpretation at our disposal we develop the required least squares algorithm, using projection update formulas.
%
Further, we present a least squares algorithm to obtain analysis filter bank coefficients in section V. Simulation
results are presented in section VI, for Gaussian as well as non-Gaussian input signals. Conclusions are presented
in section VII.
\subsection{Notations Used}
 We denote random variables by italic letters, vectors of random variable by italic bold letters and matrices of random variables with italic capital letters. Given data vectors are represented using bold letters and corresponding matrices by capital letters.  A vector $\nu$ projected on the space $S\equiv Span\{\nu_{i}\mid 1 \leq i \leq n\}$ is denoted by $\nu\mid S$ and the orthogonal complement space of S is denoted by $S^{\perp}$. $\mathbb{R}^M$ denotes the real vector space of dimension M. $\mid \mid x \mid \mid$ denotes the norm of x and it is the positive square-root of the inner-product of x with itself, where $x$ belongs to a Hilbert space.

\section{Constrained projection problem}
In this section, we first propose a lemma and observe that it implicitly defines a constrained projection problem, in a Hilbert space setting. This concept is then used, in the next section, to define the concept of signal matched whitening filter bank, which leads to optimum coding gain.

\newtheorem{lem}{Lemma}
\begin{lem}
Let $\mathcal{H}$ be a Hilbert space over $\mathbb{R}$ or $\mathbb{C}$ and let $W$ be any finite dimensional, hence closed, sub-space of $\mathcal{H}$:

$
W=Span\{\begin{array}{cccc}{\textbf {w}}_{1} & {\textbf {w}}_{2} & \cdots & {\textbf {w}}_{p}\end{array}\}
$ where, $\mathbf{w}_i \in \mathcal{H}, \forall i $.

and  

\begin{small}$ V=Span\{\begin{array}{cccc}
\mathbf{v_{1}} & 
\mathbf{v_{2}}&
\cdots &
\mathbf{v_{n}}
\end{array}\}$\end{small}, where, $\mathbf{v}_{i} \in \mathcal{H}$ for $0 \leq i \leq n$.

Let $\mathbf{\epsilon} \equiv \mathbf{x}-\sum_{i=1}^{n}{b_{i}\mathbf{v}_i}$, for any $x \in H$, where $\mathbf{\epsilon}$ is constrained to satisfy the following properties: 



1. It is confined to the orthogonal complement space of $W$, or equivalently, $\epsilon$ is orthogonal to the space $W$, and

2. $b_i$'s are chosen such that  $\mid \mid \mathbf{\epsilon} \mid \mid ^2$, i.e. square of the length of $\mathbf{\epsilon}$, is minimized, satisfying constraint 1.

Then the error, $\mathbf{\epsilon}$, admits the following representation:
  \begin{align}
\mathbf{\epsilon}=\mathbf{x}\mid W^{\perp}-\sum_{i=1}^{n}{{b_{i}}(\mathbf{v}_{i}\mid {W^{\perp}})} \label{eq:th3}
\end{align}

or equivalently:

\begin{eqnarray}
&&\hspace{-1cm}{\mathbf{\epsilon}}=\Bigg(\mathbf{x}-\sum_{k=1}^{p}{{a}_{k}\mathbf {w}_{k}}\Bigg) -\sum_{i=1}^{{n}}{{b}_{i}\Bigg({\mathbf{v}}_{i}-\sum_{k=1}^{p}{ c_{i,k} \mathbf{w}_{k}}}\Bigg)
\label{eq:theorem1}
\end{eqnarray}

\end{lem}

Proof:

From the definition of $\mathbf{\epsilon}$, we can write:
\begin{align}
\mathbf{\epsilon}=\mathbf{x}-\sum_{i=1}^{n}{{b_{i}}\mathbf{v}_{i}}\label{eq:th1}
\end{align}

Since $\mathbf{\epsilon}$ belongs to the orthogonal complement space of $W$ (from condition 1), we have:

\begin{align}
\mathbf{\epsilon}=\mathbf{\epsilon} \mid W^{\perp}\label{eq:th2}
\end{align}
  From (\ref{eq:th1}) and (\ref{eq:th2}) we get:
  \begin{align}
\mathbf{\epsilon}=\mathbf{x}\mid W^{\perp}-\sum_{i=1}^{n}{{b_{i}}(\mathbf{v}_{i}\mid {W^{\perp}})} \label{eq:th3}
\end{align}

The projections in (\ref{eq:th3}) can be written as:
\begin{align}\mathbf{x}\mid W^{\perp}=\mathbf{x}-\sum_{k=1}^{p}{{a_{k}} \mathbf{w}_{k}}\nonumber\end{align}

\begin{align} \mathbf{v}_{i}\mid W^{\perp}={{\mathbf{v}}_{i}-\sum_{k=1}^{p}{c_{i,k} \mathbf{w}_{k}}}\nonumber\end{align}
Substituting the above values in (\ref{eq:th3}) we get:

\begin{eqnarray}
&&\hspace{-1cm}{\mathbf{\epsilon}}=\Bigg(\mathbf{x}-\sum_{k=1}^{p}{{a}_{k}\mathbf {w}_{k}}\Bigg) 
 -\sum_{i=1}^{{n}}{{b}_{i}\Bigg({\mathbf{v}}_{i}-\sum_{k=1}^{p}{ c_{i,k} \mathbf{w}_{k}}}\Bigg)
\label{eq:theorem}
\end{eqnarray}
 \hspace{8cm}
   $\blacksquare$

%

Discussion:

\begin{enumerate}

%

\item In (\ref{eq:theorem}), since $b_i$'s are to be chosen such that $\mid \mid \mathbf{\epsilon}\mid \mid^2$ is minimized, $\mathbf{\epsilon}$ is essentially the error in projecting $\Big(\mathbf{x}-\sum_{k=1}^{p}{{a_{k}} \mathbf{w}_{k}}\Big)$ on the space Span$\{
\left({\mathbf{v}}_{i}-\sum_{k=1}^{p}{c_{i,k} \mathbf{w}_{k}} \mid 1 \leq i \leq n \right)\}$ .\\

\item If we specialize the result of lemma 1 to the Hilbert space of random variables, with finite mean square value, the projection in the Gaussian context is equivalent to linear combination of random variables, else it would be conditional mean. Therefore, the constrained projection problem in case of $L^2$ Gaussian random variables can be represented by (\ref{eq:theorem1}).  \\

\item The lemma 1 discusses the projection problem of (\ref{eq:th1}), where the error is constrained to lie in a subspace (here it is orthogonal complement space W), we call this problem as ``constrained projection problem". It can be very easily seen that if W is a subspace of V, then the constrained projection problem reduces to the normal unconstrained projection problem. 
\end{enumerate}
Now, in the following section, we make use of this lemma to define the concept of signal matched whitening filter bank.
\section{Signal Matched Whitening Filter Bank}
In this section we provide a definition of the concept of signal matched multirate whitening filter bank, along with the associated geometric interpretation.


\subsection{Preliminaries}

\vspace*{-1.8cm}
\begin{figure}[H]
\hspace{-0.9cm}
\includegraphics[width=0.6\textwidth, height=0.45\textwidth]{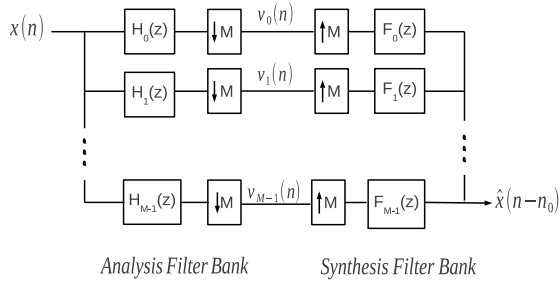}
\vspace{-3cm}
\caption{M-channel multirate filter bank} \label{fig:1}
\end{figure}

The analysis filters, $\rm {H}_i(z)=\sum_{k=0}^{N-1}{{h}_{i}(k)z^{-k}}$ for $0 \leq i \leq M-1$, shown in Fig.\ref{fig:1}, can be written in the form of M components as:
\begin{small}
\begin{eqnarray} & \hspace{-0.5cm} \rm {H}_{i}(z)=\left(\sum_{p=0}^{N/M-1}{{h}_{i}(pM)z^{-Mp}}\right)
+z^{-1}\left(\sum_{p=0}^{N/M-1}{{h}_{i}(pM+1)z^{-Mp}}\right)
\nonumber\\ & 
+ \cdots +\rm  z^{-M+1}\left(\sum_{p=0}^{N/M-1}{{h}_{i}(pM+M-1)z^{-Mp}}\right)
\end{eqnarray}\end{small}

where these M-components (written in braces), corresponding to the i-th filter, are called Type-I polyphase components and are denoted as follows: 
\begin{eqnarray}
\rm {H}_{ik}(z^M) \triangleq \sum_{p=0}^{N/M-1}{{h}_{i}(pM+k)z^{-Mp}}
\end{eqnarray}

\begin{figure}[H]
\hspace{-0.5cm}
\includegraphics[width=0.6\textwidth]{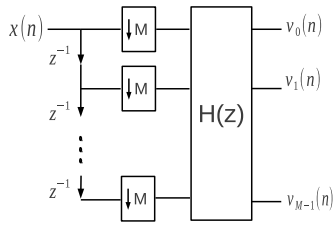}
\vspace{-110pt}
\caption{Polyphase decomposition of analysis filter bank} \label{fig:2}
\end{figure}

From the multirate filter bank theory \cite{Vaidyanathan}, using the polyphase decomposition and noble identities, the analysis filter bank can be easily re-structured to the form of single input multi output system shown in Fig. \ref{fig:2}. 

\vspace*{-1.5cm}

From fig.\ref{fig:1}, the output of i-th analysis filter can easily be observed as:
\begin{eqnarray}
{\textit{{v}}}_{i}(n)=\rm \sum_{p=0}^{{N-1}}{{{h}}_i(p)\textit{{x}}(Mn-p)}\label{eq:poly}
\end{eqnarray}

With these pre-requisite at our disposal, we are now in a position to discuss the concept of signal matched whitening filer bank and its geometrical significance.

\subsection{Geometric interpretation of Signal Matched Multirate Whitening Filter Bank}
Lets consider (\ref{eq:poly}) in more detail. If we substitute  $\rm {h}_i(p)=0$ for $0\leq p \leq M-1$, $p\neq i$ and $\rm {h}_i(i)=1$, (\ref{eq:poly}) reduces to the following form:

 \begin{eqnarray}
 {\textit{{v}}}_{i}(n)=\rm \textit{{x}}(Mn-i) +\sum_{p=M}^{(N-1)}{{{h}}_{i}(p)\textit{{x}}(Mn-p)}\label{eq:identity}
 \end{eqnarray}
 
We attach geometric significance to this expression by regarding each one of them, $0 \leq i \leq M-1$, as prediction error, therefore we can write the above equation as:
 \begin{eqnarray}
 {\textit{{e}}}_{i}(Mn-i)=\rm \textit{{x}}(Mn-i) +\sum_{p=M}^{(N-1)}{{{h}}_{i}(p)\textit{{x}}(Mn-p)}\label{eq:identity}
 \end{eqnarray}
 Clearly the channel outputs, so defined, represent different step ahead predictors starting from 1-step(when i=0) to M-step (when i=M-1), as shown in fig.\ref{fig:3}. 


\begin{figure}[H]
\hspace{1cm}
\includegraphics[width=7cm, height=6.5cm]{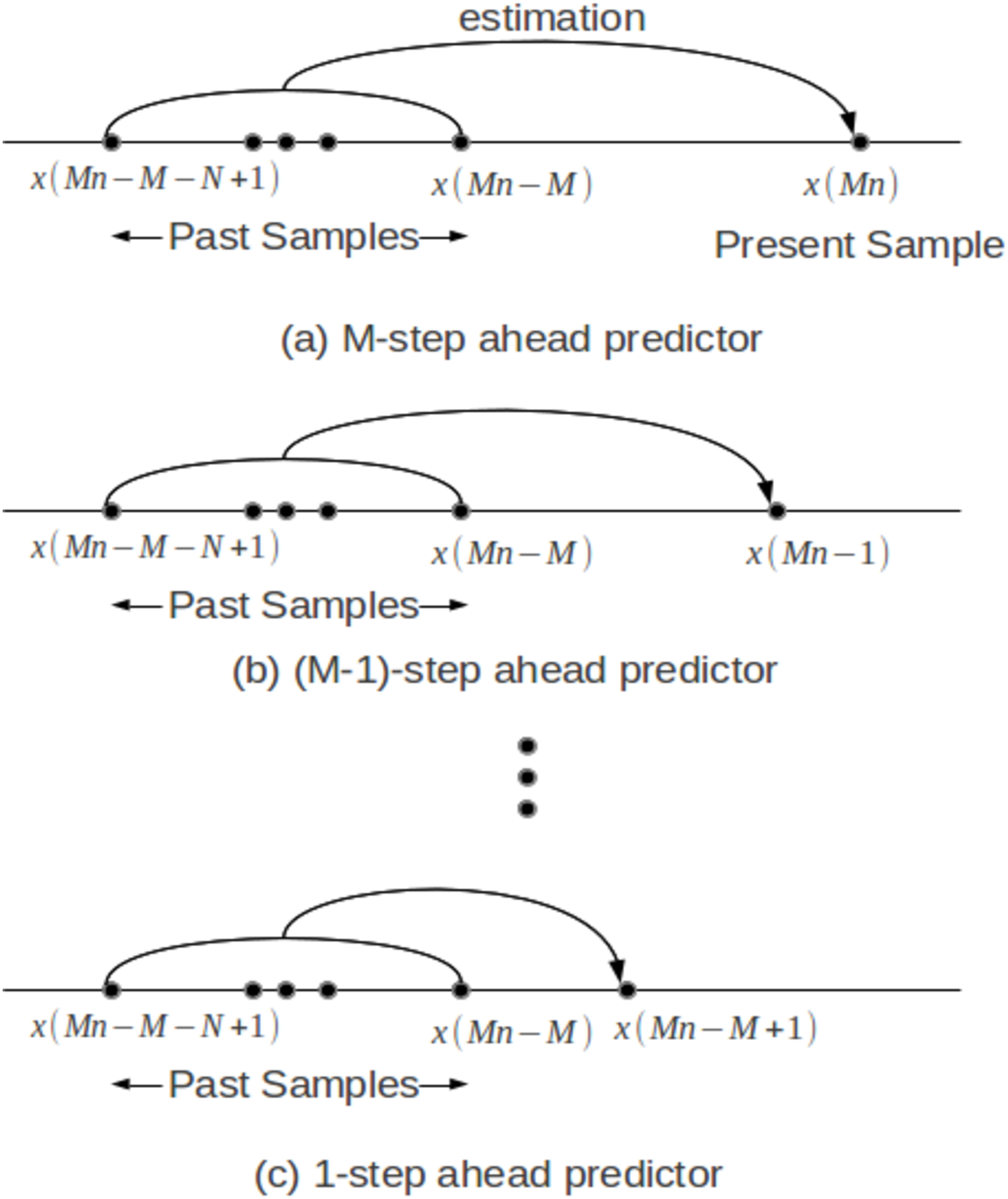}
\caption{M-forward linear predictors} \label{fig:freqres}
\end{figure}

As stated in the introduction, our objective is to achieve optimized coding gain, thus we want $\textit{e}_{i}(Mn-i)$, for $0 \leq i 
\leq M-1$, to satisfy the following conditions:\\

\begin{enumerate}[]
\item $e_{i}(Mn-i)$ should be orthogonal to space $S_{i}\equiv$Span$\{{x}(Mn-j)\mid i+1 \leq j \leq M-1\}$, for $0 \leq i \leq M-1$, to ensure orthogonalization across channels.\\

\item  $h_i$'s are chosen such that  $\mid \mid {e_{i}(Mn-i)} \mid \mid ^2$ is minimized, satisfying constraint 1, to ensure orthogonalization in time.\\
\end{enumerate}


From the above discussion and lemma 1, equation(\ref{eq:identity}) for $0 \leq i \leq M-1$ can be interpreted as a single input multi output system with outputs orthogonal in time as well as across bands, which is the requirement for maximized coding gain. If however, condition 1 is not satisfied, the channels will have some common information and thus optimized coding gain can not be obtained. Now, with all the requisite background, we are in a position to state the definition of SMWFB.
\subsection{Definition}
A single input multi output analysis filter bank, given in Fig.\ref{fig:2}, is termed as signal matched whitening filter bank (SMWFB) if the outputs $e_{i}(Mn-i)$, $0 \leq i \leq M-1$, given in (\ref{eq:identity}), satisfy the condition 1 and 2, stated above.

The conditions on (\ref{eq:identity}) have precisely the same form as mentioned in lemma 1, 
therefore $\textit{e}_{i}(Mn-i)$ can be regarded as constrained projection error and would admit the following form:

\begin{small}
\begin{eqnarray}
&&\hspace{-1cm}{\textit{e}}_{i}(Mn-i)=\Bigg(\textit{x}(Mn-i)+\sum_{k=i+1}^{M-1}{\rm g_{i}(k-i) \textit{x}(Mn-k)}\Bigg) \nonumber\\
&&\hspace{-1cm}\rm +\sum_{p=M}^{{N-1}}{{h}_{i}(p)\Bigg(\textit{{x}}(Mn-p)+}\sum_{j=i+1}^{M-1}{\rm f_{i,p}(j-i-1) \textit{x}(Mn-j)}\Bigg)\nonumber\\
\label{eq:i1}
\end{eqnarray}\end{small}
 
 or equivalently the above equation can be represented as:
 
\begin{small}
\begin{eqnarray}
& {\textit{e}}_{i}(Mn-i)=\Bigg(\textit{x}(Mn-i)-\textit{x}(Mn-i) \mid S_{i}\Bigg) \nonumber\\
& +\sum_{p=M}^{{N-1}}{\mbox{h}_{i}(p)\Bigg(\textit{{x}}(Mn-p)-\textit{{x}}(Mn-p)\mid S_{i}}\Bigg)
\label{eq:i2}
\end{eqnarray}\end{small}
 
Equation (\ref{eq:i2}) provides a precise geometrical interpretation of SMWFB as a constrained prediction problem. Equations (\ref{eq:i1}) and (\ref{eq:i2}) give the mathematical expression for SMWFB, and play a pivotal role in the understanding and the development of the required least squares algorithm.

%
%

\subsection{Observations}
In this section we discuss some observations about the SMWFB. These observations, however, are not used in the development of the required least squares algorithm, but they help in understanding the re-structuring of filter bank, as shown in Fig.\ref{fig:3} in the context of SMWFB.

Equation (\ref{eq:i1}) can be re-structured to obtain the following equation:
\begin{eqnarray}
&&\hspace{-1cm}{\textit{e}}_{i}(Mn-i)=\textit{x}(Mn-i)+ \rm \sum_{p=M}^{{N-1}}{{{h}}_{i}(p)\textit{{x}}(Mn-p)}+\nonumber
\\
&&\hspace*{1.5cm}\sum_{j=i+1}^{M-1}{\rm a_{i}(j-i-1)\textit{x}(Mn-j)}
\label{eq:i}
\end{eqnarray}

where, $\rm a_{i}(k)=\rm g_{k}(k-i-1)+\sum_{p=M}^{{N-1}}h_{i}(p){f_{i}(j-i-1)}$.
%
 
%
%
%
For $0 \leq i \leq M-1$, (\ref{eq:i}) can be written, in a vector form, as follows:

\begin{equation}
{\textit{\textbf{e}}}(Mn)=A\textit{\textbf{x}}(Mn)-\left(-\sum_{p=1}^{N/M-1}{\rm{{H}}}(p)\textit{\textbf{x}}(M(n-p))\right)\label{eq:2}
\end{equation}

where, $M \times 1$ input vector $\textit{\textbf{x}}(Mn)$ and output vector ${\textit{\textbf{e}}}(Mn)$, at time Mn, are given respectively, as:

\begin{small}\begin{eqnarray}\textit{\textbf{x}}(Mn)=\left[\begin{array}{cccccc}
\textit{x}(Mn) & \textit{x}(Mn-1) & \ldots & \textit{x}(Mn-M+1)\end{array}\right]^{T} \nonumber\end{eqnarray}\end{small}
 
\begin{small}\begin{eqnarray} {\textit{\textbf{e}}}(Mn)=\left[\begin{array}{cccccc}
\hspace*{-0.2cm}{\textit{e}}_0(Mn) \hspace*{-0.1cm} & \hspace*{-0.1cm}
 {\textit{e}}_1(Mn-1) & \hspace*{-0.2cm} \ldots \hspace*{-0.2cm}  & {\textit{e}}_{M-1}(Mn-M+1)\hspace*{-0.2cm} \end{array}\right]^{T}\nonumber \end{eqnarray} \end{small}

the matrices H(p) are given as follows:

\begin{small}
\begin{eqnarray}
\rm {{{H}}}\left(p\right)\hspace{-2pt}=\hspace{-2pt}\left[\hspace{-5pt}\begin{array}{cccccc}
 \rm{ {h}}_{0}(pM) &  \rm{{h}_{0}}(pM+1) & \ldots &  \rm{{h}_{0}}(pM+M-1)\\
 \rm{{h}_{1}}(pM) &  \rm{{h}_{1}}(pM+1) & \ldots &  \rm{{h}_{1}}(pM+M-1)\\
\vdots & \vdots & \ldots & \vdots\\
\rm{ {h}_{M-1}}(pM) &  \rm{{h}_{M-1}}(pM+1) &\ldots &  \rm{{h}_{M-1}}(pM+M-1)
\end{array}\hspace{-5pt}\right],\nonumber
\\ \hspace*{1cm} 1\leq p\leq N/M-1 \nonumber
\end{eqnarray}\end{small}


Here, 'A' is an upper triangular matrix, whose elements,
 $ a_{i}(j)'s$, are chosen in a way to orthogonalize  ${\textit{\textbf{e}}}(Mn)$.
Equation (\ref{eq:2}) implicitly leads to a structure for SMWFB, as a single input multi output system, shown in Fig. \ref{fig:3}, where A is a pre-filtering block.
\begin{figure}[H]
\vspace{-50pt}
\includegraphics[width=0.8 \textwidth]{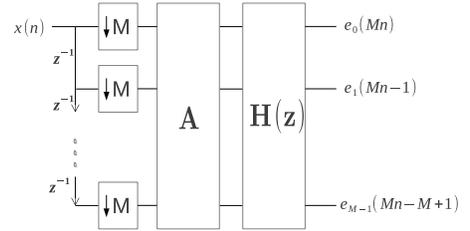}
\vspace{-160pt}
\caption{Modified signal matched analysis filter bank} \label{fig:3}
\end{figure}



 With the definition of SMWFB, provided above, we now focus on formulating the problem of least squares filtering algorithm.  
In the following section we present a Hilbert Space framework, for the given data case, and use it to present a precise problem statement.

\section{Development of the Algorithm}

\subsection{Notations and Preliminaries}
At this juncture we would like to emphasize that the lemma, presented in section II, along with its application in the present context in the form of (\ref{eq:i1}) and (\ref{eq:i2}), plays a central role in the development of the proposed algorithm. The theme of this section, essentially, is to interpret (\ref{eq:i1}) and (\ref{eq:i2}) geometrically for the given data case. For this purpose we now present the Hilbert space setting which would lead to the geometric counterpart of (\ref{eq:i1}) (or equivalently (\ref{eq:i2})).\\

\subsubsection{Notations}

We call a signal $\rm {v}(n)$ to be in a pre-windowed form if $\rm v(n)=0$ for $n < 0$. For a given discrete time signal/sequence, $\mathbf{ v}(n)$, the data vector at time `n', is defined as $1 \times L$ (L is a fixed number) vector 
\begin{small}\begin{eqnarray}\mathbf{v}(n)
\equiv \left[\begin{array}{ccccccc} 0 & \cdots & 0 & \rm v(0) & \rm v(1) & \cdots &\rm  v(n) 
\end{array}\right]\nonumber
\end{eqnarray}\end{small}
with $L \gg n$. Note that we have inserted enough number of zeros
 (the condition on L i.e. $L>>n$ ensures that the dimension, of this vector, does not change with time). This is basically a collection of all present and past values of $\rm v(n)$, upto time n, in its natural chronological order.

The corresponding  $1 \times L$ vector, for the i-th delayed and M-down-sampled version of the signal $\rm {v}(n)$ denoted as $\mathbf{v}(Mn-i)$  $\in \mathbb{R^L}$, is given as follows:

\begin{small}\begin{eqnarray}
&&\hspace*{-0.5cm}\mathbf{v}(Mn-i)\nonumber\\
&&\hspace*{-0.5cm}
\equiv \left[\begin{array}{cccccccc} 0 & \cdots & 0 & \rm v(M-i) & \rm v(2M-i) & \cdots &\rm  v(Mn-i) 
\end{array}\right] \label{eq:112}
\end{eqnarray}\end{small}

and the set of p vectors,$\{\rm v(Mn-k)|i \leq k \leq i+p-1\}$, forms a $p \times L$ matrix, denoted as $\rm V_p^{Mn-i}$, and is given as follows:

\begin{footnotesize}
\begin{align}\rm {V}_{p}^{Mn-i}\equiv
\left[\begin{array}{c}
\mathbf{\rm v}(Mn-i)\\
\mathbf{\rm v}(Mn-i-1)\\
\vdots\\
\mathbf{\rm v}(Mn-i-p+1)
\end{array}\right] \label{eq:111}\end{align}\end{footnotesize}

Here the superscript denotes the top row vector used and the subscript ``p" denotes number of rows.

The  projection operator is denoted by $\rm {P}$ and $\rm {P^{\perp}}$ denotes $\rm {(I-P)}$, the projection operator corresponding to the orthogonal complement space.

 $\rm \pi$ is the pining vector defined as $\left[\begin{array}{cccc}
 0 &  \cdots & 0 & 1
 \end{array}\right] \in \mathbb{R}^L$.\\
 
\subsubsection{Geometrical framework for the given data case}

%
Using (\ref{eq:identity}) and (\ref{eq:i1}), which defines the notion of signal matched multirate whitening filter bank and the notations defined above, we now present the geometric setting required for the development of fast least squares algorithm. Using (\ref{eq:identity}), we write all the outputs for time upto Mn, in matrix form as follows:

\begin{footnotesize}
\begin{eqnarray}
&&\left[\begin{array}{cccccc}
0 & \cdots & 0 & e_{i}(2M-i)& \cdots & e_{i}(Mn-i)\\
\end{array}\right]= \nonumber\\ &&
\hspace{.5cm}\left[\begin{array}{cccccc}
0 & \cdots & 0 & x(2M-i)& \cdots & x(Mn-i)\\
\end{array}\right]\nonumber\\
&&\hspace{.5cm}+\left[\begin{array}{cccc}
h_{i}(M) & h_{i}(M+1) & \cdots & h_{i}(M+N-1)\\
\end{array}\right]\nonumber\\
&&\hspace{.5cm}\left[\begin{array}{cccccc}
0 & \cdots & 0 & x(M) & \cdots & x(Mn-M)\\
0 & \cdots & 0 & x(M-1)& \cdots & x(Mn-M-1)\\
\vdots & \vdots &\vdots &\vdots & \cdots & \vdots \\
 0 & \cdots & 0 & x(M-N+1)& \cdots & x(Mn-M-N+1)
\end{array}\right]
\nonumber\\
\end{eqnarray}
\end{footnotesize} 
%
Using the notations, defined in the above section, the above equation can be written, in vector form, as follows:
\begin{small}
\begin{align}
&&\mathbf{{e}_{i}}^{N}(Mn-i) =\mathbf{x}(Mn-i)
 + \rm \mathbf{h}_{i}X_{N}^{Mn-M}\label{eq:8} 
\end{align}\end{small}

Since we are interested in developing an order recursive algorithm, we have introduced one additional notation, on the L.H.S. of the above equation, i.e., a super-script N, indicating the order, i.e. the number of past samples used for prediction.

Equation (\ref{eq:8}) is the `` given data case"
 counterpart of (\ref{eq:identity}) and hence can now be used to define ``given data case" version of SMWFB, discussed in section III. We, again, observe that optimized coding gain would be achieved only if the different channel outputs i.e. $\mathbf{e}_{i}(Mn-i)$,
 are orthogonal within as well as across channels. In this context we can state the conditions as follows:
 
\begin{enumerate}

\item
 $\mathbf{e}_{i}(Mn-i)$, is confined in the orthogonal space spanned by the rows of matrix 
 $\rm {X}_{M-1-i}^{Mn-i-1}$,
  i.e. 
 $S_{i}$ $\equiv Span\{\mathbf{x}(Mn-i-1),{\mathbf{x}}(Mn-i-2),\ldots{\mathbf{x}}(Mn-M+1)\}$. 
 It can be easily seen that $S_i$ is the given data case equivalent of
  $S_i$, 
  mentioned in section III. Thus projecting (\ref{eq:8}) on the orthogonal complement of $S_i$, using ({\ref{eq:Projection}}) as given in appendix, we get:

 \begin{small}\begin{eqnarray}
 &\hspace*{-0.8cm}\rm 
\mathbf{{e}_{i}^N}(Mn-i)={\mathbf{x}}(Mn-i){P^{\perp}}\left[{{X}_{M-1-i}^{Mn-i-1}}\right]
\rm +\mathbf{{h}_i}\left[{X_{N}^{Mn-M}}\right.\nonumber \\
&\hspace*{2cm}
\left.{P^{\perp}}\left[{{X}_{M-1-i}^{Mn-i-1}}\right]\right]\label{eq:9}
\end{eqnarray}\end{small}

This lead to the orthogonalization across channels, as proved in corollary 1, given in appendix, for given data case.

\item $h_i$'s are chosen such that  $\mid \mid \mathbf{e}_{i}(Mn-i)\mid \mid ^2$ is minimized, satisfying constraint 1.
\end{enumerate}

It is then obvious that (\ref{eq:8}), under these conditions become precisely the constrained prediction problem stated in lemma 1. With the above two conditions and equation(\ref{eq:Projection}), 
$\mathbf{e}_{i}(Mn-i)$  
can easily be written as:

\begin{small}\begin{eqnarray}\rm 
\mathbf{{e}^{N}_{i}}(Mn-i)={\mathbf{x}}(Mn-i){P^{\perp}}\left[{{X}_{M-1-i}^{Mn-i-1}}\right]
\rm {P^{\perp}}\left[{X_{N}^{Mn-M}}\right.
\nonumber\\
\hspace{2cm}\left.{P^{\perp}}\left[{{X}_{M-1-i}^{Mn-i-1}}\right]\right],
\label{eq:10}
\end{eqnarray}\end{small}
where $0 \leq i \leq M-1$.

Equation (\ref{eq:10}) is precisely the geometric counterpart of (\ref{eq:i1}), for the given data case, which will now be used to develop the required least squares algorithm. The errors $\mathbf{e}_{i}(Mn-i)$'s, so obtained would be orthogonal across time as well as across various channels. 

\subsection{Problem Statement}
With (\ref{eq:10}) at our disposal, the problem statement of least squares whitening filter bank can be stated as follows:
``Given a set
of pre-windowed data samples $\{\rm{x}(k),0\leq k\leq Mn\}$ at time Mn, of a zero mean, wide sense stationary
process, obtain a single input multi output system, as given in (\ref{eq:10}), which takes present data. i.e. $x(Mn)$ as input and produces M-whitened outputs orthogonal in time as well as across bands, using a time and order recursive exact least squares algorithm."

%
%
\subsection{Development of the Algorithm}

We now have the required geometrical framework for the given data case problem, to develop the least squares algorithm.
Here, our main objective is to obtain the outputs, $e_{i}(Mn-i)$'s, at the present instant, hence we post-multiply both sides of (\ref{eq:10}) with transpose
of pinning vector, $\mathbf{\pi}$, and also replace fixed order N by a variable p:

\begin{small}
\begin{eqnarray}\rm 
{e}_{i}^{p}(Mn-i)={\mathbf{x}}(Mn-i){P^{\perp}}\left[{{X}_{M-1-i}^{Mn-i-1}}\right]
\rm P^{\perp}\left[{X_{p}^{Mn-M}}
{P^{\perp}}\left[{{X}_{M-1-i}^{Mn-i-1}}\right]\right]\pi^{T}\nonumber\\
\label{eq:11}
\end{eqnarray}
\end{small}

The complete set of recursions, to compute time and order updates of $\rm {e}_{i}^{p}(Mn-i)$'s, can be easily obtained by proper substitutions in the inner product update formula \cite{Friedlander}. For a quick reference the inner-product update relation is given in appendix as (equation (\ref{eq:inner_product_formula})). All the auxiliary quantities arising in the algorithm are first defined in table I and their correlations are defined in table II. The substitutions required to obtain the recursions are given in table III. In order to give an idea about how the entire algorithm is developed, we provide some sample computations and then present the algorithm in table IV.


The auxiliary quantities are defined by substituting the entries of table I in the following equation:

\begin{equation}
\mathbf{e}=\mathbf{y}\rm P^{\perp}[Y]\mathbf{w^{T}}\label{eq:16}
\end{equation}
where $\mathbf{e}$ is the error vector involved in projecting $\mathbf{y}$ onto the space spanned by the rows of the matrix $\rm{Y}$, see (\ref{eq:Projection}) in appendix. 

Example:
By substituting the entries of fifth row of table I, in (\ref{eq:16}),
%
 we obtain the definition of backward prediction error of i-th channel, given as follows:

\begin{align}
&{r}_{i}^{p}(Mn-M)=\mathbf{x}(Mn-M-p-1)\rm P^{\perp} \left[{X}_{M-1-i}^{Mn-i-1}\right]\nonumber\\
&\rm P^{\perp}\left[X_{p}^{Mn-M}
\rm P^{\perp} \left[{X}_{M-1-i}^{Mn-i-1}\right]\right]\mathbf{\pi}^{T}
\label{eq:r}
\end{align}

Example: Substitute the entries of second row of table I in (\ref{eq:16}),
%
 to obtain:

\begin{align}
&&{\gamma}^{q}(n)=\mathbf{x}(n-i-q-1)\rm P^{\perp}\left[{X}_{q}^{n-1}\right]\mathbf{\pi}^{T}
\label{eq:gamma}
\end{align}
 
%
The order update relations, for prediction error and auxiliary quantities, are calculated by substituting the entries from table \ref{table:substitution} in the inner product update formula \cite{Friedlander}, (except for the order updates of $\mathbf{r}^{p}_{i}(Mn-M)$ and $ \delta_{i}^{p}(Mn-M),\:0\leq i\leq M-1$, which is discussed latter).  

Example : Substitute the values from second row of table III, in the inner product update formula (\ref{eq:inner_product_formula}) to compute the order update for $\rm e_{i}^{p}(Mn-i)$:\\
we get:

\begin{align}\rm 
{e}^{p+1}_{i}(Mn-i)={e}^{p}_{i}(Mn-i)-\Delta^p_{{e}_i,{r}_i}(Mn-i)\nonumber\\
\rm .R^{-{r}_i}_{p}(Mn-M).{r}_i^{p}(Mn-M) 
\label{eq:frwd_error_update_relation}
\end{align}

We illustrate the time update computations in this algorithm with the following example:

Example 4: Substitute the entries of third row of table III in (\ref{eq:corollary}):
we get:

\begin{align}
R_{p}^{{r}_{i}}(kM-M)=R_{p}^{{r}_i}((k-1)M-M)+{r}_{i}^{p}(kM-M).\nonumber\\
\delta_{p}^{-i}(kM-M){r}_{i}^{p}(kM-M)
\end{align}

Discussion:

Since, the order update of $\mathbf{r}^{p}_{i}(Mn-M)$ and $ \delta_{i}^{p}(Mn-M),\:0\leq i\leq M-1$ are not computed using the inner-product update formula rather using a update relation, proposed in \cite{joshiI}, for quick reference it is given in appendix as equation(\ref{eq:corollary}). To illustrate this further, we present a few examples:
%
%
 Time and order update of $r_{M-1}^{p}(Mn-M)$, are computed using a joint process estimator. From (\ref{eq:r}), for the specific case of i=M-1, and (\ref{eq:gamma}) it can easily be noted that $\rm {r_{M-1}^{p}}(Mn-M)$ is $\mathbf{\gamma_{p}}(n)$
down sampled with M and delayed by 1. 
  Updates of backward residual error
for rest of multirate filter bank 
are calculated using the updates of $\rm r_{M-1}^{p}(Mn-M)$ and a corollary proposed in \cite{joshiII}, for reference this update relation is given in appendix (refer equation (\ref{eq:corollary})). As an example let us substitute the following values in (\ref{eq:corollary})

\begin{small} $\rm \mathbf{\nu}={\mathbf{x}}(Mn-M-p-1){P^{\perp}}\left[{{X}_{M-i-1}^{Mn-i-1}}\right]$,
 $\rm x={\mathbf{x}}(Mn-i-1){P^{\perp}}\left[{{X}_{M-i-2}^{Mn-i-1}}\right]$, $\rm {V_{1:n}={X_{p}^{Mn-M}}{P^{\perp}}\left[{{X}_{M-i-1}^{Mn-i-1}}\right]}$ and  $\rm w=\pi$
\end{small}
we get:

\begin{align}\rm 
& r_{i}^{p}(Mn-M)=\rm {\mathbf{x}}(Mn-M-p-1){P^{\perp}}\left[{{X}_{M-i-1}^{Mn-i-1}}\right]\nonumber\\
&\rm {P^{\perp}}\left[{X_{p}^{Mn-M}}{P^{\perp}}\left[{{X}_{M-i-1}^{Mn-i-1}}\right]\right]\mathbf{\pi}^{T}\nonumber\\
&=\rm r_{i+1}^{p}(Mn-M)-\Delta_{r_{i+1},{e}_{i+1}}^{p}(Mn-M)\nonumber\\
&\rm .R_{p}^{-{e}_{i+1}}(n).{e}_{i+1}^{p}(Mn-i)
\label{eq:15} 
\end{align}

Similarly, all the order and time updates are computed, using substitution values from table \ref{table:substitution}, and brought together to give
the LS algorithm, given in table \ref{table:LS_algo}. The recursions of the algorithm give rise to a lattice-ladder structure as shown in Fig.\ref{fig:analysis}.

\subsection{Complexity}
The number of arithmetic operations required at every time step to compute the outputs of the above algorithm are:  (7+6M)N+7M additions and (14+12M)N+14M multiplications, where N is the order of the filters and M is the number of channels.


\renewcommand{\arraystretch}{1.4}
\newcounter{temptablecounter1}

\begin{figure*}
\vspace{2cm}
\setcounter{temptablecounter1}{\value{figure}}
\begin{table}[H]
\begin{scriptsize}

\caption{Definition of the auxiliary quantities and parameters, for $0\leq i\leq M-1$.}
\label{table:def_aux}
\begin{tabular}{|lm{2.7cm}m{2.7cm}m{2.7cm}m{2.7cm}|}
\hline 
\multicolumn{1}{|l}{e} & \multicolumn{1}{l}{$\nu$} & \multicolumn{1}{l}{U} & \multicolumn{1}{l}{$w$} & \multicolumn{1}{l|}{$\Theta$}  \tabularnewline
\hline

for $0 \leq q \leq M-1-i $ &  &  &  &  \tabularnewline

${{\epsilon}}_i^{q}(Mn-i)$ 
& ${\mathbf{x}}(Mn-i)
$ & ${{X}_{q}^{Mn-i-1}}
$ & $\pi $ 
& $\mathbf{{\hat{a}}^{i}_{q}} =\left[\begin{array}{cccc}
{\hat{a}}^{i}_{1} & {\hat{a}}^{i}_{2} & \ldots & {\hat{a}}^{i}_{q}\end{array}\right]$

\tabularnewline

${{\gamma}}_i^{q}(Mn-i-1)$ & ${\mathbf{x}}(Mn-i-q-1)
$ & ${{X}_{q}^{Mn-i-1}}
$ & $\mathbf{\pi}$
& $\mathbf{{\hat{b}}^{i}_{q}}=\left[\begin{array}{cccc}
{\hat{b}}^{i}_{1} & {\hat{b}}^{i}_{2} & \ldots & {\hat{b}}^{i}_{q}\end{array}\right]$

\tabularnewline

$\widehat{\delta}_{q}^{i}(Mn-i-1)$ & $\mathbf{\pi}$ & $ \rm {{X}_{q}^{Mn-i-1}}
$ & $\mathbf{\pi}$
& -------
\tabularnewline 
 
 ${e}{i}^{{p}}(Mn-i)$ & $\begin{aligned}{\mathbf{x}}(Mn-i)\\
\rm P^{\perp} \left[{{X}_{M-1-i}^{Mn-i-1}}\right]
\end{aligned}
$ & $\begin{aligned}{X_{p}^{Mn-M}}\\
\rm P^{\perp} \left[{{X}_{M-1-i}^{Mn-i-1}}\right]
\end{aligned}
$ & $\pi \rm P^{\perp} \left[{{X}_{M-1-i}^{Mn-i-1}}\right]$
&  $\mathbf{{h}^{i}_{p}}$$=\left[\begin{array}{cccc}
{h}^{i}_{1} & {h}^{i}_{2} & \ldots & {h}^{i}_{p}\end{array}\right]$

\tabularnewline

${r}_{i}^{p}(Mn-M)$ & $\begin{aligned}{\mathbf{x}}(Mn-M-p-1)\\
\rm P^{\perp} \left[{{X}_{M-1-i}^{Mn-i-1}}\right]
\end{aligned}
$ & $\begin{aligned}{X_{p}^{Mn-M}}\\
\rm P^{\perp} \left[{{X}_{M-1-i}^{Mn-i-1}}\right]
\end{aligned}
$ & $\pi \rm P^{\perp} \left[{{X}_{M-1-i}^{Mn-i-1}}\right]$ 
&$\mathbf{g^{i}_{p}}$$=\left[\begin{array}{cccc}
g^{i}_{1} & g^{i}_{2} & \ldots & g^{i}_{p}\end{array}\right]$

\tabularnewline

$\delta_{p}^{i}(Mn-M)$ & $\mathbf{\pi} \rm P^{\perp} \left[{{X}_{M-1-i}^{Mn-i-1}}\right]$ & $\begin{aligned}{X_{p}^{Mn-M}}\\
\rm P^{\perp}\left[{{X}_{M-1-i}^{Mn-i-1}}\right]
\end{aligned}
$ & $\mathbf{\pi} \rm P^{\perp}\left[{{X}_{M-1-i}^{Mn-i-1}}\right]$
&   -------
\tabularnewline 

$\mathbf{e^{p}}(n)$ & $\mathbf{x}\left(n\right)$ & ${X_{p}^{(n-1)}}$ & $\mathbf{\pi}$ 
& $\mathbf{c_{p}}=\left[\begin{array}{cccc}
c_{1} & c_{2} & \ldots & c_{p}\end{array}\right]$

\tabularnewline

$\mathbf{r}^{p}\left(n-1\right)$ & $\mathbf{x}\left(n-p-1\right)$ & ${X_{p}^{(n-1)}}$ & $\mathbf{\pi}$ 
& $\mathbf{d_{p}}$$=\left[\begin{array}{cccc}
d_{1} & d_{2} & \ldots & d_{p}\end{array}\right]$
\tabularnewline
\hline 
\end{tabular}
\end{scriptsize}
\end{table}
\setcounter{figure}{\value{temptablecounter1}}
\end{figure*}

\newcounter{temptablecounter2}
\begin{figure*}
\vspace{-0.5cm}
\setcounter{temptablecounter2}{\value{figure}}

\hspace{2.5cm}
\begin{table}[H]
\begin{footnotesize}\caption{Autocorrelation and cross-correlation coefficients}
\label{table:corr}

\begin{tabular}{|lll|}
\hline 
$\nu$ & $w$ & $\nu w^{T}$\tabularnewline
\hline 

$\mathbf{{\epsilon}}_i^{q}(Mn-i)$ & $\mathbf{{\epsilon}}_i^{q}(Mn-i)$ & $R_{q}^{{\epsilon}_i}(Mn-i)$
\tabularnewline

$\mathbf{{\gamma}}^{q}_i(Mn-i-1)$ & $\mathbf{{\gamma}}^{q}_i(Mn-i-1)$ & $R_{q}^{{\gamma}_i}(Mn-i-1)$
\tabularnewline

$\mathbf{{\epsilon}}^{q}_i(Mn-i)$ & $\mathbf{{\gamma}}^{q}_i(Mn-i-1)$ & $\Delta_{{\epsilon_i},{\gamma}_i}^{q}(Mn-i)$\tabularnewline
$\mathbf{{\gamma}}^{q}_i(Mn-i-1)$ & $\mathbf{{\epsilon}}_i^{q}(Mn-i)$ & $\Delta_{{\gamma}_i,{\epsilon_i}}^{q}(Mn-i-1)
$\tabularnewline

$\mathbf{{e}}_{i}^{p}(Mn-i)$ & $\mathbf{{e}}_{i}^{p}(Mn-i)$ & $R_{p}^{{e}_{i}}(Mn-i)$
\tabularnewline

$\mathbf{{r}}_{i}^{p}(Mn-M)$ & $\mathbf{{r}}_{i}^{p}(Mn-M)$ & $R_{p}^{{r}_{i}}(Mn-M)$
\tabularnewline

$\mathbf{{e}}_{i}^{{p}}(Mn-i)$ & $\mathbf{{r}}_{i}^{p}(Mn-M)$ & $\Delta_{{e}_{i},{r}_{i}}^{p}(Mn-i)$
\tabularnewline

$\mathbf{{r}}_{i}^{p}(Mn-M)$ & $\mathbf{{e}_{i}^{{p}}}(Mn-i)$ & $\Delta_{{r}_{i},{e}_{i}}^{p}(Mn-M)$\tabularnewline

$\mathbf{{e}^{p}}(n)$ & $\mathbf{{e}^{p}}(n)$ & $R_{p}^{{e}}\left(n\right)$\tabularnewline

$\mathbf{r}^{p}(n-1)$ & $\mathbf{r}^{p}(n-1)$ & $R_{p}^{r}(n-1)$\tabularnewline

$\mathbf{e}^{p}(n)$ & $\mathbf{r}^{p}(n-1)$ & $\Delta_{e,r}^{p}(n)$\tabularnewline

$\mathbf{r}^{p}(n-1)$ & $\mathbf{e}^{p}(n)$ & $\Delta_{r,e}^{p}(n-1)$\tabularnewline

\hline 
\end{tabular}\end{footnotesize}

\end{table}
\setcounter{figure}{\value{temptablecounter2}}
\vspace*{-4pt}
\end{figure*}

\newcounter{temptablecounter3}
\begin{figure*}
\renewcommand\arraystretch{2.5} 
\setcounter{temptablecounter3}{\value{figure}}
\begin{table}[H]\vspace{0.5cm}

\begin{scriptsize}
\caption{Substitution table for least squares algorithm }
\label{table:substitution}
\begin{tabular}{|llllllll|}
\hline 
\multicolumn{1}{|l}{$\begin{aligned}Recursion\\ \hspace*{-1cm} No.\end{aligned}$ } & \multicolumn{1}{l}{$\mathbf{\nu}$} & \multicolumn{1}{l}{$\rm {V_{1:p}}$} & \multicolumn{1}{l}{$\mathbf{x}$} & \multicolumn{1}{l}{$\mathbf{w}$} & \multicolumn{1}{l}{$\mathbf{\nu_{p+1}}$} & $\begin{aligned}\mathbf{\nu}\rm P^{\perp}\left[\mathbf{x}\right]\rm P^{\perp}\left[{V_{1:p+1}}\rm P^{\perp}\left[\mathbf{x}\right]\right]\\
\mathbf{w^{T}}\end{aligned}$ &\multicolumn{1}{l|}{$\begin{aligned}Update\\relation\end{aligned}$}\tabularnewline
\hline

(1) & $\mathbf{x}\left(n\right)$ & $\rm {X_{p}^{(n-1)}}$ & $\mathbf{\pi}$ & $\mathbf{x}\left(n-p-1\right)$ & 0 & $\Delta_{e,r}^{p}(n-1)$ & (\ref{eq:inner_product_formula}) \tabularnewline
 
(2) & $\mathbf{x}\left(n-p-1\right)$ & $\rm {X_{p}^{(n-1)}}$ & $\mathbf{\pi}$ & $\mathbf{x}\left(n\right)$ & 0 & $\Delta_{r,e}^{p}(n-2)$ & (\ref{eq:inner_product_formula})  \tabularnewline

(3) & $\mathbf{x}\left(n-p-1\right)$ & $\rm {X_{p}^{(n-1)}}$ & $\mathbf{\pi}$ & $\mathbf{x}\left(n-p-1\right)$ & 0 & $R_{p}^{r}\left(n-2\right)$ & (\ref{eq:inner_product_formula}) \tabularnewline

(4) & $\mathbf{x}\left(n\right)$ & $\rm {X_{p}^{(n-1)}}$ & $\mathbf{\pi}$ & $\mathbf{x}\left(n\right)$ & 0 & $R_{p}^{e}\left(n-1\right)$ & (\ref{eq:inner_product_formula})  \tabularnewline
 
(5) & $\mathbf{\pi}$ & $\rm {X_{p}^{(n-1)}}$ & 0 & $\mathbf{\pi}$ & $\mathbf{x}\left(n-p-1\right)$ & $\delta_{p+1}(k-1)$ & (\ref{eq:inner_product_formula})  \tabularnewline 
 
(6) & $\mathbf{x}\left(n\right)$ & $\rm {X_{p}^{(n-1)}}$ & 0 & $\mathbf{\pi}$ & $\mathbf{x}\left(n-p-1\right)$ & $e^{p+1}\left(n\right)$ & (\ref{eq:inner_product_formula}) \tabularnewline
 
(7) & $\mathbf{x}\left(n-p-1\right)$ & $\rm {X_{p}^{(n-1)}}$ & 0 & $\mathbf{\pi}$ & $\mathbf{x}\left(n\right)$ & $r^{p+1}\left(n\right)$ & (\ref{eq:inner_product_formula}) \tabularnewline

(8) & ${\mathbf{x}}(Mn-i)$ & $\left[{{X}_{q}^{Mn-i-1}}\right]$ & $\mathbf{\pi}$ & $\mathbf{x}(Mn-i-q-1)$ & 0 & $\Delta_{{\epsilon},{\gamma}}^{q}(Mn-i)$ & (\ref{eq:inner_product_formula}) \tabularnewline

(9) & $\mathbf{x}(Mn-i-q-1)$ & $\left[{{X}_{q}^{Mn-i-1}}\right]$ & $\mathbf{\pi}$ & ${\mathbf{x}}(Mn-i)$ & 0 & $\Delta_{{\gamma},{\epsilon}}^{q}(Mn-M+1)$ & (\ref{eq:inner_product_formula}) \tabularnewline

(10) & $\mathbf{x}(Mn-i-q-1)$ & $\left[{{X}_{q}^{Mn-i-1}}\right]$ & $\mathbf{\pi}$ & $\mathbf{x}(Mn-i-q-1)$ & 0 & $R_{q}^{\gamma}(Mn-M+1)$ & (\ref{eq:inner_product_formula}) \tabularnewline

(11) & $\begin{aligned}{\mathbf{x}}(Mn-i)\\
\rm P^{\perp}\left[{{X}_{M-1-i}^{Mn-1-i}}\right]
\end{aligned}
$ & $\begin{aligned}{X_{p}^{Mn-M}}\\
\rm P^{\perp}\left[{{X}_{M-1-i}^{Mn-1-i}}\right]
\end{aligned}
$ & $\pi \rm P^{\perp}\left[{{X}_{M-1-i}^{Mn-1-i}}\right]$ & $\begin{aligned}{\mathbf{x}}(Mn-i)\\
\rm P^{\perp}\left[{{X}_{M-1-i}^{Mn-1-i}}\right]
\end{aligned}
$ & 0 & $R_p^{{\epsilon}_q}(Mn-i)$
& (\ref{eq:inner_product_formula}) \tabularnewline

(13) & $\mathbf{x}(Mn-i-q-1)
$ & $\left[{{X}_{q}^{Mn-i}}\right]
$ & 0 & $\pi$ & ${\mathbf{x}}(Mn-i)
$ & $\gamma^{q+1}(Mk-i)
$ & (\ref{eq:inner_product_formula}) \tabularnewline

(14) & ${\mathbf{x}}(Mn-i)
$ & $\left[{{X}_{q}^{Mn-i-1}}\right]
$ & 0 & $\pi$ & $\mathbf{x}(Mn-i-q-1)
$ & ${{\epsilon}}^{q}(Mn-i)
$ & (\ref{eq:inner_product_formula}) \tabularnewline

(15) & $\begin{aligned}{\mathbf{x}}(Mn-M-p-1)\\{P^{\perp}}\left[{{X}_{M-i-1}^{Mn-i-1}}\right]\end{aligned}$ & 

$\begin{aligned}{X_{p}^{Mn-M}}\\{P^{\perp}}\left[{{X}_{M-i-1}^{Mn-i-1}}\right]\end{aligned}$ & 

 $\begin{aligned}{\mathbf{x}}(Mn-i-1)\\{P^{\perp}}\left[{{X}_{M-i-2}^{Mn-i-1}}\right]\end{aligned}$ & $\mathbf{\pi}$ & 0 & $ r_{i}^{p}(Mn-M)$ & (\ref{eq:corollary})\tabularnewline

(16) & $\begin{aligned}{\mathbf{x}}(Mn-M-p-1)\\
\rm P^{\perp}\left[{{X}_{M-1-i}^{Mn-1-i}}\right]
\end{aligned}
$ & $\begin{aligned}{X_{p}^{Mn-M}}\\
\rm P^{\perp}\left[{{X}_{M-1-i}^{Mn-1-i}}\right]
\end{aligned}
$ & $\pi \rm P^{\perp}\left[{{X}_{M-1-i}^{Mn-1-i}}\right]$ & $\begin{aligned}{\mathbf{x}}(Mn-M-p-1)\\
\rm P^{\perp}\left[{{X}_{M-1-i}^{Mn-1-i}}\right]
\end{aligned}
$ & 0 & $R_p^{{r}_i}(Mn-M)$ & (\ref{eq:inner_product_formula}) \tabularnewline

(17) & $\begin{aligned}{\mathbf{x}}(Mn-i)\\
\rm P^{\perp}\left[{{X}_{M-1-i}^{Mn-1-i}}\right]
\end{aligned}
$ & $\begin{aligned}{X_{p}^{Mn-M}}\\
\rm P^{\perp}\left[{{X}_{M-1-i}^{Mn-1-i}}\right]
\end{aligned}
$ & $\pi \rm P^{\perp}\left[{{X}_{M-1-i}^{Mn-1-i}}\right]$ & $\begin{aligned}{\mathbf{x}}(Mn-i)\\
\rm P^{\perp} \left[{{X}_{M-1-i}^{Mn-1-i}}\right]
\end{aligned}
$ & 0 & $R_p^{e_i}(Mn-i)$ & (\ref{eq:inner_product_formula}) \tabularnewline
%

(18) & $\begin{aligned}{\mathbf{x}}(Mn-i)\\
\rm P^{\perp}\left[{{X}_{M-1-i}^{Mn-1-i}}\right]
\end{aligned}
$ & $\begin{aligned}{X_{p}^{Mn-M}}\\
\rm P^{\perp}\left[{{X}_{M-1-i}^{Mn-1-i}}\right]
\end{aligned}
$ & $\pi \rm P^{\perp}\left[{{X}_{M-1-i}^{Mn-1-i}}\right]$ & $\begin{aligned}{\mathbf{x}}(Mn-M-p-1)\\
\rm P^{\perp}\left[{{X}_{M-1-i}^{Mn-1-i}}\right]
\end{aligned}
$ & 0 & $\Delta_{e_{i},{r}_{i}}^{p}(Mn-i)$ & (\ref{eq:inner_product_formula}) \tabularnewline

(19) & ${\mathbf{\pi}}{P^{\perp}}\left[{{X}_{M-i-1}^{Mn-i-1}}\right]$ & 
$\begin{aligned}{X_{p}^{Mn-M}}\nonumber\\{P^{\perp}}\left[{{X}_{M-i-1}^{Mn-i-1}}\right]\end{aligned}$ & 
 $\begin{aligned}{\mathbf{x}}(Mn-i-1)\\{P^{\perp}}\left[{{X}_{M-i-2}^{Mn-i-1}}\right]\end{aligned}$ & $\mathbf{\pi}$ & 0 & $\delta_{p}^{i}(kM-M)$ & (\ref{eq:corollary})\tabularnewline

(20) & $\begin{aligned}{\mathbf{x}}(Mn-i)\\
\rm P^{\perp}\left[{{X}_{M-1-i}^{Mn-i-1}}\right]
\end{aligned}
$ & $\begin{aligned}{X_{p}^{Mn-M}}\\
\rm P^{\perp}\left[{{X}_{M-1-i}^{Mn-i-1}}\right]
\end{aligned}
$ & 0
 & $\mathbf{\pi\rm P^{\perp}}
 \left[{{X}_{M-1-i}^{Mn-i-1}}\right]$
  & $\begin{aligned}{\mathbf{x}}(Mn-M-p-1)\\
\rm P^{\perp}\left[{{X}_{M-1-i}^{Mn-1-i}}\right]
\end{aligned}
$ & $ e^{p+1}_{i}(Mn-i)$ & (\ref{eq:inner_product_formula})  \tabularnewline

\hline 
\end{tabular}
\end{scriptsize}
\end{table}
\setcounter{figure}{\value{temptablecounter3}}
\end{figure*}

\renewcommand{\arraystretch}{2}
\newcounter{temptablecounter4}
\begin{figure*}
\vspace{1cm}
\setcounter{temptablecounter4}{\value{figure}}
\begin{table}[H]
\begin {center}\begin{scriptsize}\caption{The LS algorithm for the analysis side}
\label{table:LS_algo}
\begin{tabular}{|ll|}
\hline 
$\begin{aligned} & Set\, all\,\Delta s\, and\, Rs\, to\,0\, at\, k=0\\
 & \rm e^{0}(n)=r^{0}(k)=x(k)\,\,\,\\
 & \rm {\epsilon}^0(Mk-i)={\gamma}^0(Mk-i)=x(Mk-i)\\
 & \rm {r}_{M-1}^{0}(k-1)=r_{0}(kM-M)\\
 & For\, p=1\, to\, N,\, i=0\, to\, M-1,k=0\, to\, n, 
 where\, N\, is\, order\, of\, the\, filter\, ,\, n\, is\, latest\, time\, and\, M\, is \, number\, of\, channel\,
\end{aligned}
$ & \tabularnewline
\hline 
$\Delta_{e,r}^{p}(k)=\Delta_{e,r}^{p}(k-1)+e^{p}(k)$.$\delta_{p}^{^{-1}}(k-1).r^{p}(k-1)$ & (1)\tabularnewline
$\Delta_{r,e}^{p}(k-1)=\Delta_{r,e}^{p}(k-2)+r^{p}(k-1).\delta_{p}^{-1}(k-1).e^{p}(k)$ & (2) \tabularnewline
$R_{p}^{r}\left(k-1\right)=R_{p}^{r}(k-2)+r^{p}(k-1).\delta_{p}^{-1}(k-1).r^{p}(k-1)$& (3) \tabularnewline
$R_{p}^{e}\left(k\right)=R_{p}^{e}(k-1)+e^{p}(k).\delta_{p}^{-1}(k-1).e^{p}(k)$ & (4) \tabularnewline
$\delta_{p+1}(k-1)=\delta_{p}(k-1)-r^{p}(k-1).R_{p}^{-r}(k-1).r^{p}(k-1)$ & (5) \tabularnewline
$e^{p+1}(k)=e^{p}(k)-\Delta_{e,r}^{p}(k).R_{p}^{-r}(k-1)r^{p}(k-1)$ & (6) \tabularnewline
$r^{p+1}\left(k\right)=r^{p}(k-1)-\Delta_{r,e}^{p}(k-1)R_{p}^{-e}(k).e^{p}(k)$ & (7) \tabularnewline
\hline

for i=M-1 to 0; & \tabularnewline
$\Delta_{{\epsilon},{\gamma}}^{q}(Mk-i)=\Delta_{{\epsilon},{\gamma}}^{q}(M(k-1)-i)+{\epsilon}^{q}(Mk-i)$.$\hat{\delta}_{q}^{^{-1}}(Mk-i-1).
{\gamma}^{q}(Mk-i-1)$ & (8) \tabularnewline
$\Delta_{{\gamma},{\epsilon}}^{q}(Mk-i-1)=\Delta_{{\gamma},\tilde{\epsilon}}^{q}(M(k-1)-i-1)+\gamma^{q}(Mk-i-1).\widehat{\delta}_{q}^{-1}(Mk-i-1).\tilde{\epsilon}^{q}(Mk-i)$ & (9) \tabularnewline
$R_{q}^{\gamma}(Mk-i-1)=R_{q}^{\gamma}(M(k-1)-i-1)+\gamma^{q}(Mk-i-1).\widehat{\delta}_{q}^{-1}(Mk-i-1).\gamma^{q}(Mk-i-1)$ & (10) \tabularnewline
$R_{q}^{{\epsilon}}(Mk-i)=R_{q}^{{\epsilon}}(M(k-1)-i)+{\epsilon}^{q}(Mk-i).\widehat{\delta}_{q}^{-1}(Mk-i-1).R_{q}^{{\epsilon}}(Mk-i)$ & (11) \tabularnewline
$\widehat{\delta}_{q+1}^{-1}(Mk-i-1)=\widehat{\delta}_{q}^{-1}(Mk-i-1))-\gamma^{q}(Mk-i-1).R_{q}^{\gamma}(Mk-i-1).\gamma^{q}(Mk-i-1)$ & (12) \tabularnewline
${\epsilon}^{q+1}(Mk-i)={\epsilon}^{q}(Mk-i)-\Delta_{{\epsilon},\gamma}^{q}(Mk-i).R_{q}^{-\gamma}(Mk-i-1).\gamma^{q}(Mk-i-1)$ & (13) \tabularnewline
$\gamma^{q+1}(Mk-i)=\gamma^{q}(Mk-i-1)-\Delta_{\gamma,{\epsilon}}^{q}(Mk-i-1).R_{q}^{-{\epsilon}}(Mk-i).{\epsilon}^{q}(Mk-i)$ & (14) \tabularnewline
\hline 

$e^{0}_{i}(Mk-i)={\epsilon}^{M-1-i}(Mk-i)$  & \tabularnewline
for $i<M-1$ & \tabularnewline
$r_{i}^{p}(kM-M)=r_{i-1}^{p}(kM-M)-\Delta_{{r}_{i-1},{e}_{i-1}}^{p}(kM-M).R_{p}^{-{e}_{i-1}}(kM).{e}_{i-1}^{p}(kM)$ & (15) \tabularnewline
\hline 
$R_{p}^{{r}_{i}}(kM-M)=R_{p}^{{r}_i}(kM-2M)+{r}_{i}^{p}(kM-M).\delta_{p}^{-i}(kM-M){r}_{i}^{p}(kM-M)$  & (16) \tabularnewline
$\begin{aligned}R_{p}^{e_{i}}(kM)=R_{p}^{e_{i}}(kM-M)+{e}_{i}^{p}(kM).\delta_{p}^{-i}(kM-M).{e}_{i}^{p}(kM)\end{aligned}
$  & (17) \tabularnewline
$\Delta_{e_{i},{r}_{i}}^{p}(kM)=\Delta_{e_{i},{r}_{i}}^{p}(kM-M)+{e}_{i}^{p}(Mk).\delta_{p}^{-i}(Mk-M).{r}_{i}^{p}(Mk-M)$ & (18) \tabularnewline
$\delta_{p}^{i}(kM-M)=\delta_{p}^{i}(kM-M)-{r}_{i}^{p}(kM-M).R_{p}^{-{r}_{p}}(Mk-M).{r}_{i}^{p}(Mk-M)$ & (19) \tabularnewline
${e}_{i}^{{p+1}}(Mk)={e}_{i}^{p}(Mk)-\Delta_{{e}_{i},r_i}^{p}(Mk).R_{p}^{-{r}_{i}}(kM).{r}_i^{{p}}(Mk)$ & (20) \tabularnewline
\hline 
\end{tabular}
\end{scriptsize}\end{center}
\end{table}
\setcounter{figure}{\value{temptablecounter4}}
\end{figure*}

\newcounter{tempfigcounter1}

\renewcommand{\arraystretch}{1.4}

\begin{figure*}
\vspace{-0.05cm}
\setcounter{temptablecounter2}{\value{figure}}
\begin{tiny}
\begin{table}[H]
\caption{Substitution table for parameter estimation }
\label{table:substitution_par_est}
\begin{tabular}{|lllll|}
\hline 
\multicolumn{1}{|l}{z} & \multicolumn{1}{l}{$V_{1:n-1}/V_{2:n}$} & \multicolumn{1}{l}{$x$} & \multicolumn{1}{l}{$\nu_{n}/\nu_{1}$}  & \multicolumn{1}{l|}{Update Equation no.} \tabularnewline
\hline 
$\mathbf {x}\left(k\right)$ & $\rm X_{p}^{k-1}$ & 0 & $\mathbf{x}(k-p-1)$ & (1)\tabularnewline
 
$\mathbf{x}\left(k-p-1\right)$ & $\rm X_{p}^{k-1}$ & 0 & $\mathbf{x}(k)$ & (2) and (7)\tabularnewline

$\begin{aligned}\mathbf{x}(kM-i)
\end{aligned}
$ & $\begin{aligned}\rm X_{p}^{Mn-i-1}
\end{aligned}
$ & 0 & $\begin{aligned}\mathbf{x}(Mk-i-p-2)
\end{aligned}
$ & (3) \tabularnewline
$\begin{aligned}\mathbf{x}(Mk-i-p-2)
\end{aligned}
$ & $\begin{aligned}\rm X_{p}^{Mn-i-1}
\end{aligned}
$ & 0 & $\begin{aligned}\mathbf{x}(Mk-i) 
\end{aligned}
$ & (4)\tabularnewline
$\begin{aligned}\mathbf{x}(kM-M-p-1)\\
\\
\end{aligned}
$ & $\begin{aligned}\rm X_{p}^{Mk-M}
P^{\perp}\left[{X}_{M-2-i}^{Mk-i-2}\right]
\end{aligned}
$ & 0 & $\begin{aligned}\mathbf{x}(Mk-M-p-1)\rm 
P^{\perp}\left[{X}_{M-2-i}^{Mk-M+1}\right]
\end{aligned}
$ & (5)\tabularnewline
$\begin{aligned}\mathbf{x}(Mk-i)\\
\\
\end{aligned}
$ & $\begin{aligned}\rm X_{p}^{Mk-M}
P^{\perp}\left[{X}_{M-1-i}^{Mk-i-1}\right]
\end{aligned}
$ & 0 & $\begin{aligned}\mathbf{x}(Mk-M-p-1) \rm 
P^{\perp}\left[{X}_{M-1-i}^{Mk-M+1}\right]
\end{aligned}
$ & (6)\tabularnewline
\hline 
\end{tabular}
\end{table}\end{tiny}
\setcounter{figure}{\value{temptablecounter2}}
\end{figure*}

\begin{figure*}
\setcounter{temptablecounter3}{\value{figure}}
\begin{table}[H]
\caption{The LS algorithm for parameter estimation}
\label{table:LS_par_est}
\begin{tabular}{|ll|}
\hline 
$\begin{aligned} & Initialize
 & c_0(k)=d_{0}(k)=\hat{a}_i^0(k)=\hat{b}_i^0(k)={h}_i^0(k)={g}_i^0(k)=0 \end{aligned}$  & \tabularnewline
\hline 
$ \mathbf{c_{p+1}}(k)=
\left[\mathbf{c_{p}}(k)\mid0\right]-\Delta_{e,r}^{p}(k).R_{p}^{-r}(k-1)\left[\mathbf{d_{p}(k)}\mid1\right]$ & {1}\tabularnewline
$\mathbf{d_{p+1}}\left(k\right)$$=\left[\mathbf{0\mid d_{p}}(k-1)\right]-\Delta_{r,e}^{p}(k-1)R_{p}^{-e}(k).\left[\mathbf{1\mid c_{p}(k)}\right]$ & {2}\tabularnewline

$ \mathbf{\hat{a}_{p+1}}(k)=
\left[\mathbf{\hat{a}_{p}}(k)\mid0\right]-\Delta_{{\epsilon},\gamma}^{q}(Mk-i).R_{q}^{-\gamma}(Mk-M+1).\left[\mathbf{\hat{b}_{p}(k)}\mid1\right]$ & {3} 
\tabularnewline
$\mathbf{\hat{b}}_{p+1}\left(k\right)=\left[\mathbf{0\mid \hat{b}}(k-1)\right]-\Delta_{\gamma,{\epsilon}}^{q}(Mk-M+1).R_{q}^{-{\epsilon}}(Mk-i).\left[\mathbf{1\mid \hat{a}_{p}(k)}\right]$ & {4}
\tabularnewline

for $i<M-1$  & \tabularnewline
$\mathbf{g_{i}^{p+1}}(kM)$$=\left[\mathbf{0\mid g_{i}^{p}}(kM-M)\right]-\Delta_{{r}_{i-1},{e}{i-1}}^{p}(kM-M).R_{p}^{-e_{i-1}}(kM-M).\left[\mathbf{1\mid{h}_{i}^{p}(k-1)}\right]$ & {5}\tabularnewline

%
$\begin{aligned}\mathbf{{h}_{i}^{p+1}}(kM)=\left[\mathbf{{h}_{i}^{p}}(kM)\mid0\right]-\Delta_{{\epsilon},\gamma}^{q}(Mk-i).R_{q}^{-\gamma}(Mk-M+1).[\mathbf{a_i^{M-1-i}}]-\\
\Delta_{e_{i},{r}_{i}}^{p}(kM).R_{p}^{-{r}_{i}}(kM-M).\left[\mathbf{g_{i}^{p}(kM)}\mid1\right]\end{aligned}$ & {6}\tabularnewline
 
$\mathbf{g_{M-1}^{p+1}}(kM-M)=\mathbf{d_{p+1}}\left(kM-M\right)$ & {7}\tabularnewline
$\begin{aligned}
\mathbf{a}_i^{p}=-\left(\mathbf{x}(Mn-i)+\rm {h}_i^{N}X_{p}^{Mn-M}\right)
[{{X}_{M-1-i}^{Mn-i-1}}]^{T}
\left[\rm {{X}_{M-1-i}^{Mn-i-1}} \left[\rm{{X}_{M-1-i}^{Mn-i-1}}\right]^{T}\right]^{-1}
\end{aligned}$
& {8}
 \tabularnewline
\hline 
\end{tabular}
\end{table}
\setcounter{figure}{\value{temptablecounter3}}
\end{figure*}

\begin{figure*}
\setcounter{tempfigcounter1}{5}
\begin{figure}[H]
\vspace{-10pt}
\includegraphics[width=14cm, height=6cm]{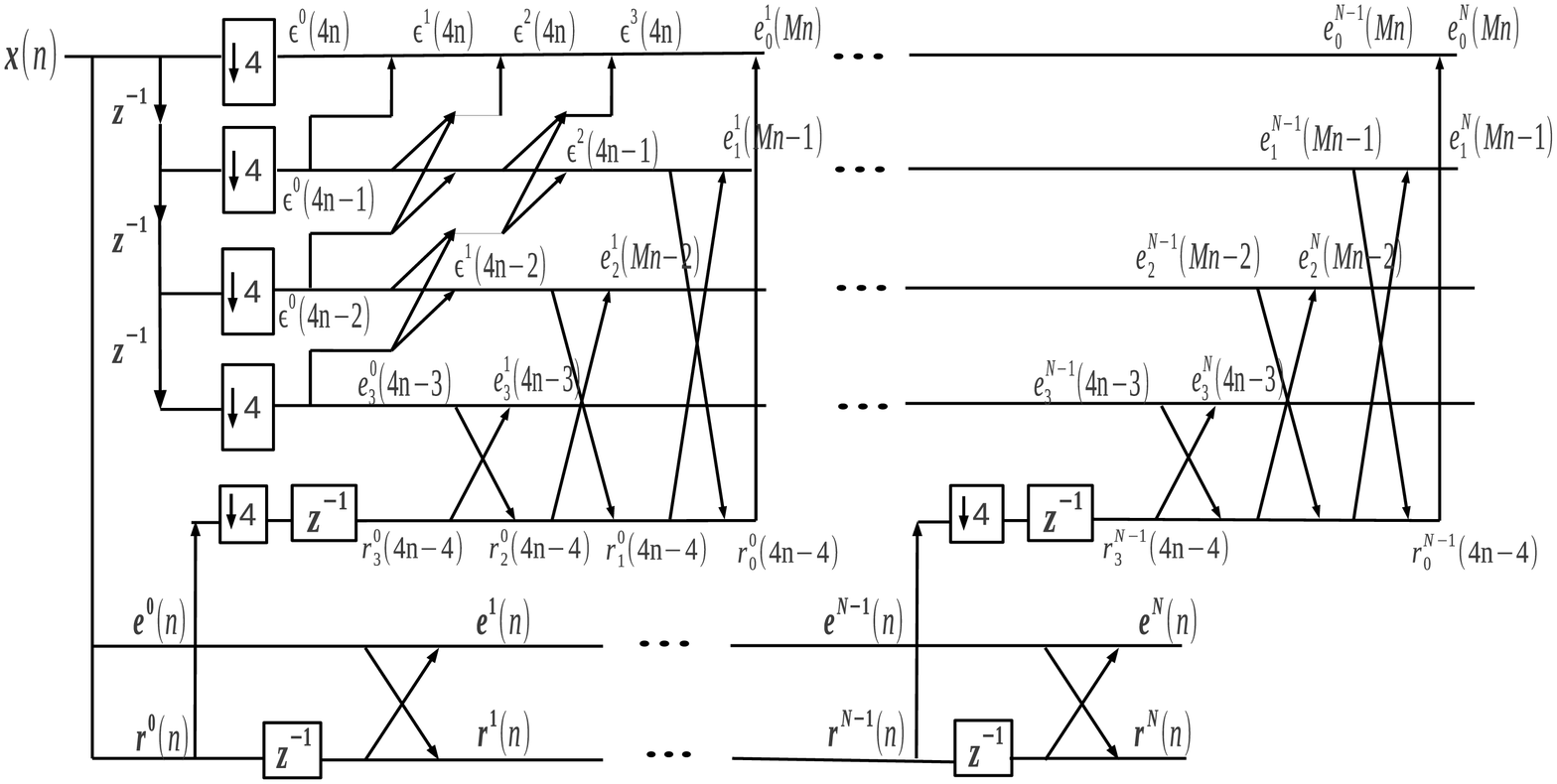}

\hspace{2cm}
\caption{Lattice-ladder structure for the proposed Analysis filter bank}\label{fig:analysis}
\end{figure}
\setcounter{figure}{\value{tempfigcounter1}}
\end{figure*}

\section{Least Squares Estimation of analysis filter bank}

In the previous section, least squares algorithm
 for signal matched whitening filter bank is discussed. Since the algorithm gives rise to a lattice-like structure, and not a direct form of filter bank, the filter bank
 coefficients are not directly available. We now present a least squares algorithm to compute these coefficients, i.e. the 
$a_{i}'s$ and ${h}_{i}'s$ for $0\leq i\leq M-1$
 as given in (\ref{eq:i}),
 using the lattice parameters. The required
  ${h}_{i}'s$ 
  are the least squares estimates of the parameters, and can be obtained using (\ref{eq:least_squres_par_est}) as given in appendix, as:

\begin{small}
\begin{align*} &\rm
\mathbf{{h}_{i}^{p}} =-\mathbf{x}(Mn-i)\rm P^{\perp}\left[{{X}_{M-1-i}^{Mn-i-1}}\right]
 \left[\rm P^{\perp}\left[{{X}_{M-1-i}^{Mn-i-1}}\right]\left[{X_{p}^{Mn-M}}\right]^{T}\right]\nonumber\\ \rm
& \hspace*{1.2cm}\left[{X_{p}^{Mn-M}}\rm P^{\perp}\left[{{X}_{M-1-i}^{Mn-i-1}}\right]\left[{X_{p}^{Mn-M}}\right]^{T}\right]^{-1}
\end{align*}
\end{small}
\begin{small}$\mathbf{{h}_{i}^{p}}$\end{small} 
here, \\
\begin{small}$
\rm P^{\perp}\left[{{X}_{M-1-i}^{Mn-i-1}}\right]
\left[{X_{p}^{Mn-M}}\right]^{T}
 \rm [{X_{p}^{Mn-M}}\rm P^{\perp}
 \left[{{X}_{M-1-i}^{Mn-i-1}}
 \right]\left[{X_{p}^{Mn-M}}\right]^{T}
 ]^{-1}$\end{small} is the generalized inverse of \begin{small}$\begin{small}\rm {X_{p}^{Mn-M}}
 \rm P^{\perp}\left[{{X}_{M-1-i}^{Mn-i-1}}\right]\end{small}$.\end{small}

%

 It can be observed that in order to calculate the order update recursions for filter bank parameters, we need an update relation for the generalized inverse.
In \cite{joshiII} an update relation for a similar space has been proposed, for a quick reference it is given in appendix, as equation(\ref{eq:pseudo_inverse_update}). Now, all the required recursions can be obtained with proper substitutions in this update relation. And the various auxiliary quantities arising in the recursions are defined in table \ref{table:def_aux}.
As we proceeded in section IV, to illustrate how the algorithm is developed, some recursions are discussed and then we present the algorithm in table VI.

Example 1: Definition of auxiliary quantities:

Substituting the entries of fifth row in (\ref{eq:least_squres_par_est}) we get the least squares estimate of $\mathbf{{g}_{i}^{p}}$, which is given as:
\begin{small}
\begin{align*} &\rm
\mathbf{{g}_{i}^{p}} =-\mathbf{x}(Mn-M-p-1)\rm P^{\perp}\left[{{X}_{M-1-i}^{Mn-i-1}}\right]
 \left[\rm P^{\perp}\left[{{X}_{M-1-i}^{Mn-i-1}}\right]\left[{X_{p}^{Mn-M}}\right]^{T}\right]\nonumber\\ \rm
& \hspace*{1.2cm}\left[{X_{p}^{Mn-M}}\rm P^{\perp}\left[{{X}_{M-1-i}^{Mn-i-1}}\right]\left[{X_{p}^{Mn-M}}\right]^{T}\right]^{-1}
\end{align*}
\end{small}

Example 2: Update recursions:

To obtain the order update recursions for \begin{small}$\mathbf{{h}_{i}^{p+1}}$ \end{small}, substituting entries of sixth row of table V in (\ref{eq:pseudo_inverse_update}), i.e.:

\begin{small}$\mathbf{z}=\mathbf{x}(Mn-i)$;  $\rm V_{1:n-1}=\rm X_{p}^{Mn-M}P^{\perp}\left[{X}_{M-1-i}^{Mn-i-1}\right]$\end{small}
 \begin{small}$\mathbf{\nu}_{n}=\mathbf{x}(Mn-M-p-1)\rm P^{\perp}\left[{X}_{M-1-i}^{Mn-i-1}\right]$\end{small} and \begin{small}
 $\mathbf{x}=\rm X^{Mn-1-i}_{M-1-i}$
 \end{small}.\\
We get:
\begin{small}
\begin{eqnarray}
  &{\mathbf{h}}_{i}^{p+1}(Mn-i) = \left[{\mathbf{h}}_{i}^{p}(Mn-i) \mid0 \right]- 
  \Delta_{{\epsilon},\gamma}^{q}(Mk-i)\nonumber\\
  &.R_{q}^{-\gamma}(Mk-M+1){\mathbf{\hat{a}}}_{i}^{M-1-i}(Mn-i)\nonumber\\
 &\hspace*{1cm} -\Delta_{{\epsilon}_{i},r^{i}}^{p}(Mn-i).R_{p}^{-r^{i}}(Mn-M).\left[\mathbf{g}_{i}^{p}(Mn-M)\mid1\right]\nonumber\\
 &\label{eq:18}
\end{eqnarray}
\end{small}

From the above expression we can observe the relationship between the parameter vector of the filter bank and the lattice coefficients. 



 Least squares estimate of $\mathbf{a_i}'s$ can be computed from (\ref{eq:9}), as given below: 
\begin{small}\begin{align}
&\mathbf{a}_i^{p}=-\left(\mathbf{x}(Mn-i)+\rm {h}_i^{N}X_{p}^{Mn-M}\right)
[{{X}_{M-1-i}^{Mn-i-1}}]^{T}\nonumber\\
&\left[\rm{{X}_{M-1-i}^{Mn-i-1}} \left(\rm{{X}_{M-1-i}^{Mn-i-1}}\right)^{T}\right]^{-1}
\end{align}\end{small}

 Recursive update relation for $\mathbf{a_i}'s$ is not calculated here, however by using matrix inversion lemma  computational complexity is reduced.

\subsection{Complexity}
In order to compute $\mathbf{h_i}$ for $0 \leq i \leq M-1$, $7N^2$ additions and $7N^2$ multiplications are required, however to compute $\mathbf{a_i}$ for $0 \leq i \leq M-1$, using matrix inversion lemma, need $6N^{2}+N$ operations, as shown in \cite{joshiII}.

\section{Simulation}
We now present simulation results to validate the proposed algorithm. Since we have not come across any other work dealing with adaptive algorithm in the context of signal adapted multirate filter banks, we compare our work with most recent block processing algorithms in the context.

\subsection{Coding gain}In order to benchmark the results obtained here, we compare coding gain obtained using the proposed algorithm with those present in the literature. We will first state the coding gain definition used, here, to obtain the results. 

A generalized coding gain expression for sub-band coders was given by Katto et al. in \cite{Katto} and later a simplified solution of the same was presented in \cite{Aase} by Aase et al., which is as follows:
\begin{eqnarray}
G_{SBC}=\dfrac{\sigma_{x_q}^2}{\sigma_{e_q}^2}
\end{eqnarray}
where,
${\sigma_{x_q}^2}$ is the input variance and 
$\sigma_{e_q}^2$ is the average quantized output variance, as discussed in \cite{PLT}, it can be written as:
 \begin{eqnarray}
 \sigma_{e_q}^2=\dfrac{1}{M}\sum_{0}^{M-1}{\sigma_{e_i}^2} 
 \end{eqnarray}
 where, 
 $\sigma_{e_i}^2$ is the error variance of the unquantized i-th channel output of the proposed analysis filter bank,
 $b_i$ is the bit rate of i-th channel and the average bit rate b.
%
 For maximum coding gain $\sigma_{e_q}^2$ should be minimum, using the arithmetic mean geometric mean inequality, which is:
  \begin{eqnarray}
 \sigma_{e_q}^2 \geq c2^{-2b_i}\left(\prod_{i=0}^{M-1}{\sigma_{e_i}^2}\right)^{1/M} \label{eq:condition_am_gm}
 \end{eqnarray}
 In case of equality for the above relation, minimum $\sigma_{e_q}^2$ is obtained and this is possible when all the error variances are same. It will be proved using simulations, this condition is approximately obtained using the proposed algorithm. 
 
 Therefore, the coding gain can be written as:
 
 \begin{eqnarray}
G_{SBC}=\dfrac{{\sigma_{x}}^2}{(\prod_{i=0}^{M-1}{\sigma_{e_i}^2})^{1/M}}
\end{eqnarray}

Experiment 1: Simulations are performed for a four channel SMWFB, for a filter length 8. We consider two cases of AR(2) input, with poles at $0.975e^{\pm j\theta}$ and compare the coding gain of proposed filter bank with the results presented by Lu et. al. \cite{Lu}, in table \ref{Table:cg_compared_Lu}.\\

\begin{table}[H]
\begin{center}\begin{small}
\caption{Coding Gain Comparison}
\label{Table:cg_compared_Lu}
\begin{tabular}{lll}
\hline 
 \multicolumn{1}{c}{$\theta$}  & \multicolumn{1}{c}{ Biothogonal Filter Bank, \cite{Lu}} & \multicolumn{1}{c}{Modified SMFB}\tabularnewline
\hline 

 $\pi/2.8$ & 6.6411 & 14.5564  \tabularnewline

 $\pi/1.75$ &   4.9174 & 10.5967  \tabularnewline
\hline 

\end{tabular}
\end{small}\end{center}
\end{table}

\begin{figure}[H]
        \includegraphics[width=8cm, height=3cm]{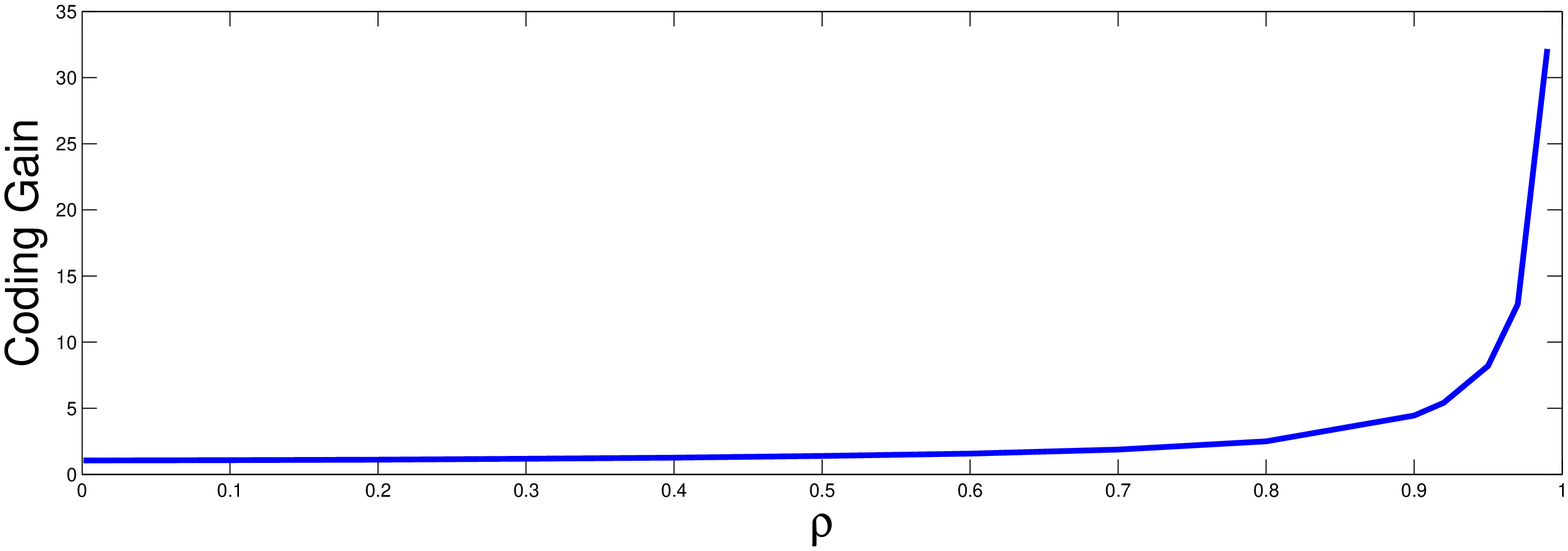}
        \caption{Coding gain as $\rho$ goes towards unity}
        \label{fig:pole_position}
    \end{figure}

Experiment 2: The response of the proposed filter bank to an AR(2) filter with poles at ${\rho}e^{\pm j\pi/3}$ as $\rho$ moves towards unit circle is shown in Fig.{\ref{fig:pole_position}}\\

Experiment 3: In \cite{weng}, design of a GTD-biorthogonal sub-band coder is presented, it is shown that this filter bank offers better coding gain 
as compared to existing sub-band coders. As number of channels increases the coding gain comes closer to the theoretically 
maximum coding gain. Here, 
for the same input signal, an AR(2) process with poles at $ 0.975e^{\pm j{\pi}/3}$, we show that the coding gain converges when number of channel is as small as 2, this result is embedded in 
table \ref{table:cg_different_M}. The filter bank is designed with each analysis filter of order 5 and number of channels are 4.\\


\begin{table}[h!]
\begin{center}\begin{small}
\caption{Behavior of Coding gain (in dB) as number of channels change}
\label{table:cg_different_M}
\begin{tabular}{|l|l|l|}
\hline 
\multicolumn{1}{|c|}{Number } & \multicolumn{1}{c|}{$\begin{aligned}G_{SBC}\\
\end{aligned}$
using }
 & \multicolumn{1}{c|}{Approximate $G_{SBC}$ }
\tabularnewline

of channels & the proposed algorithm &from \cite{weng}
\tabularnewline
\hline 
$2$ & 11.7872 & 10.4\tabularnewline
\hline 
$3$ & 11.4975 & 9.9 \tabularnewline
\hline
$4$ & 11.8174 & 11.2 \tabularnewline
\hline
$5$ & 11.8492 & 11.3\tabularnewline
\hline  
$6$ & 11.8081 & 11.1\tabularnewline
\hline   
\end{tabular}
\end{small}\end{center}
\end{table}

\subsection{Convergence of filter bank parameters}
\begin{figure}[H]
    \begin{subfigure}[b]{0.5\textwidth}
        \includegraphics[width=8cm, height=3cm]{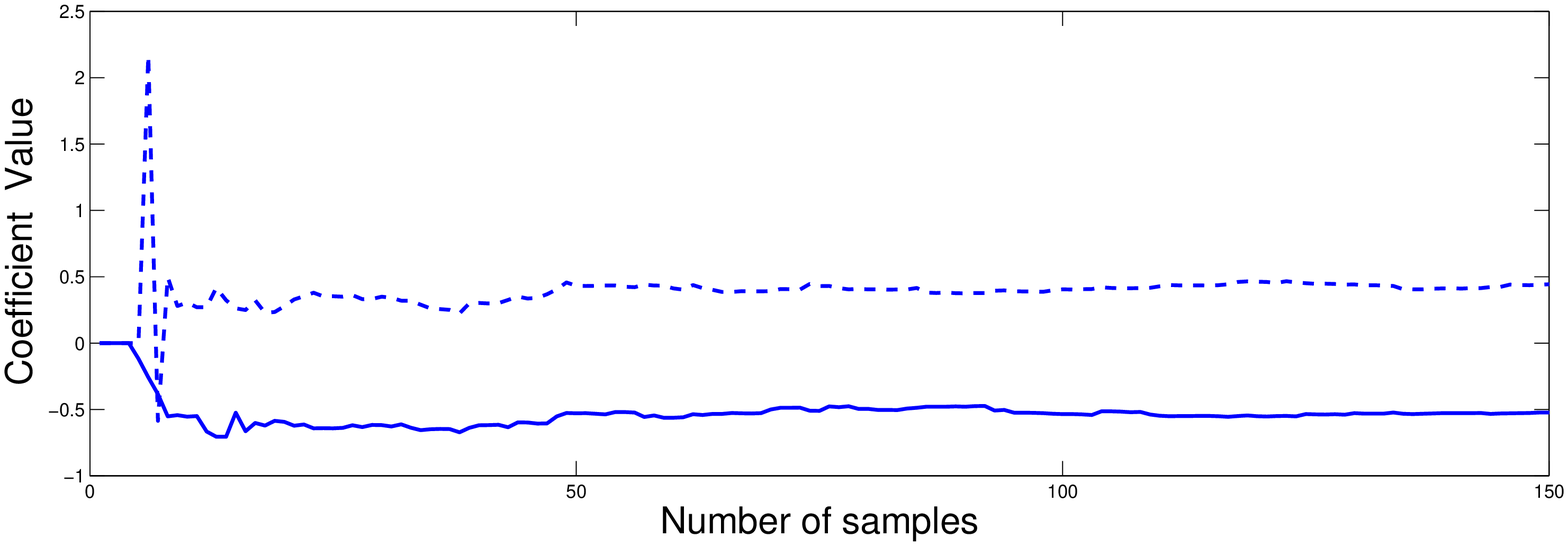}
         \caption{}
    \end{subfigure}   
    \begin{subfigure}[b]{0.5\textwidth}
        \includegraphics[width=8cm, height=3cm]{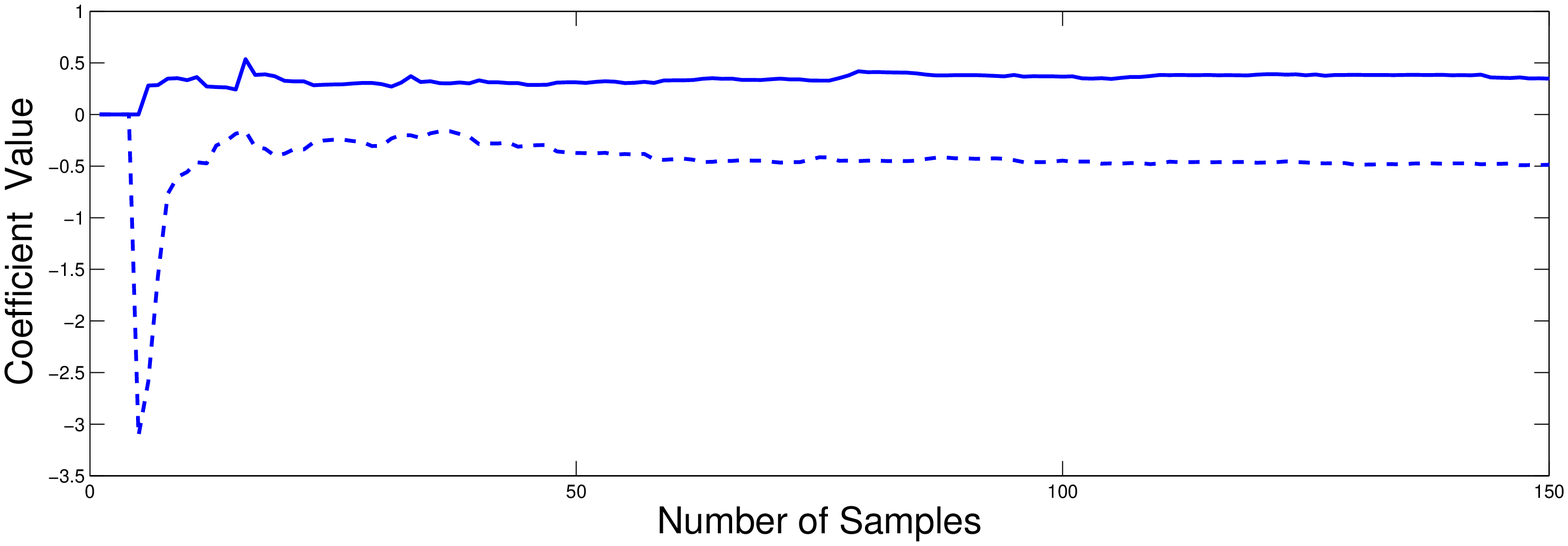}
        \caption{}
    \end{subfigure}
    \caption{For input signal of Experiment 2(i), Parameter trajectory of filter coefficients (a)$\rm H_{0}$ and (b)$\rm H_{1}$.}
    \label{fig:convergence_propoerty(i)}
\end{figure}

We illustrate the convergence property of the proposed algorithm for estimation of filter bank parameters, using the following examples:

Experiment 4:
(i) Here the input signal is an AR(2) input, with poles at $0.6e^{\pm j\pi/3}$, number of bands are 2 and filter length is 3, the convergence of the filter parameters as time progresses is shown in Fig.{\ref{fig:convergence_propoerty(i)}}. These parameters are computed using the algorithm given in section III and can be interpreted using (\ref{eq:2}). In part(a) of Fig.{\ref{fig:convergence_propoerty(i)}}, the bold line represents $h_{0}(1)$ and dashed line gives the trajectory of $h_{0}(2)$ and in part(b) the bold line is for $h_{1}(1)$ and the dashed line for $h_{1}(2)$. In all the following examples of Experiment 4, these representations are followed.\\

\begin{figure}[H]
    \begin{subfigure}[b]{0.5\textwidth}
        \includegraphics[width=8cm, height=3cm]{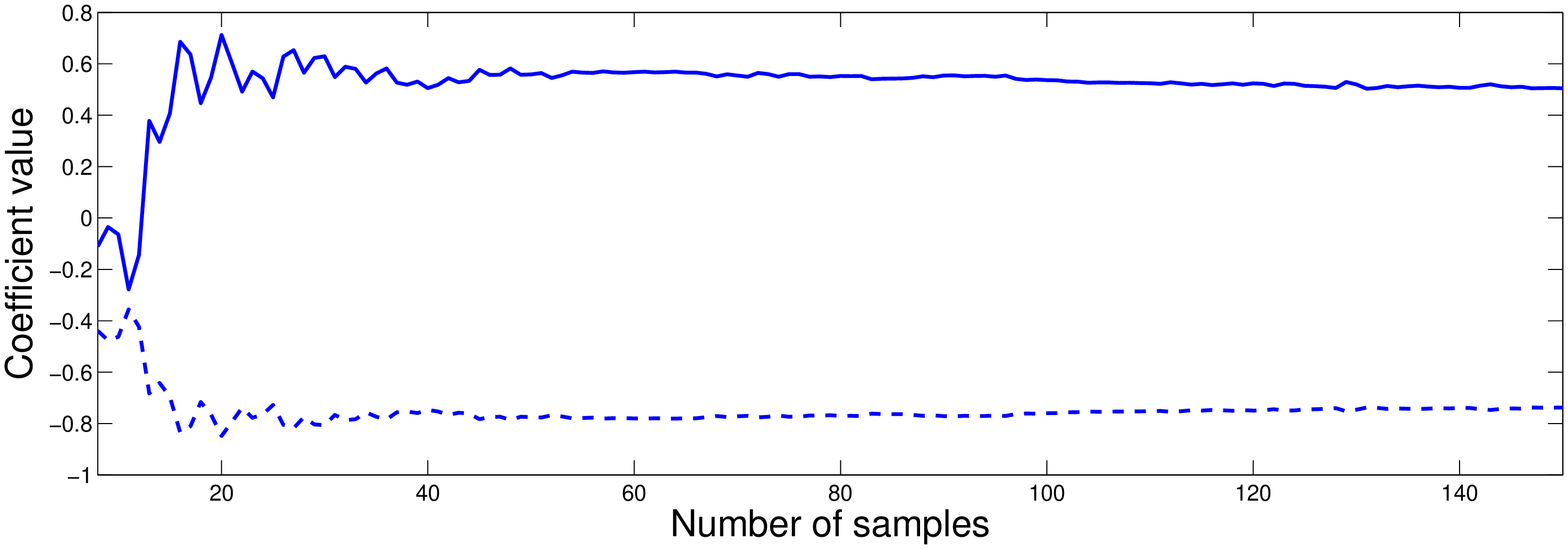}
         \caption{}
    \end{subfigure}   
    \begin{subfigure}[b]{0.5\textwidth}
        \includegraphics[width=8cm, height=3cm]{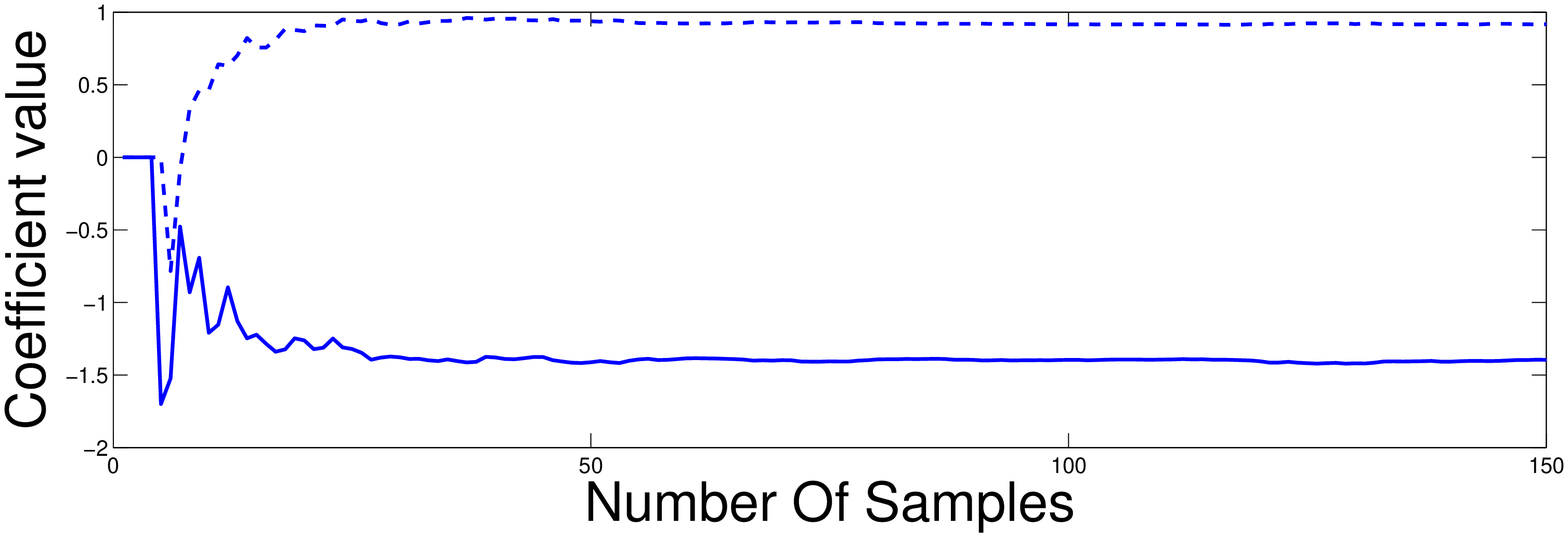}
        \caption{}
    \end{subfigure}
    \caption{For input signal of Experiment 2(ii), Parameter trajectory of filter coefficients (a)$\rm H_{0}$ and (b)$\rm H_{1}$.}
    \label{fig:convergence_propoerty(ii)}
\end{figure}

(ii) Here the input signal is an AR(2) input, with poles at $0.9e^{\pm j\pi/3}$, number of bands are 2 and filter length is 3, the convergence of the filter parameters as time progresses is shown in Fig.{\ref{fig:convergence_propoerty(ii)}}.\\

(iii) Here the input signal is an AR(2) input, with poles at $0.9e^{\pm j\pi/3}$, number of bands are 2 and filter length is 3, same as above but this AR filter has been excited using a white input with gamma distribution. The convergence of the filter parameters as time progresses is shown in Fig.{\ref{fig:convergence_propoerty(iii)}}.\\

\begin{figure}[H]
    \begin{subfigure}[b]{0.5\textwidth}
        \includegraphics[width=8cm, height=3cm]{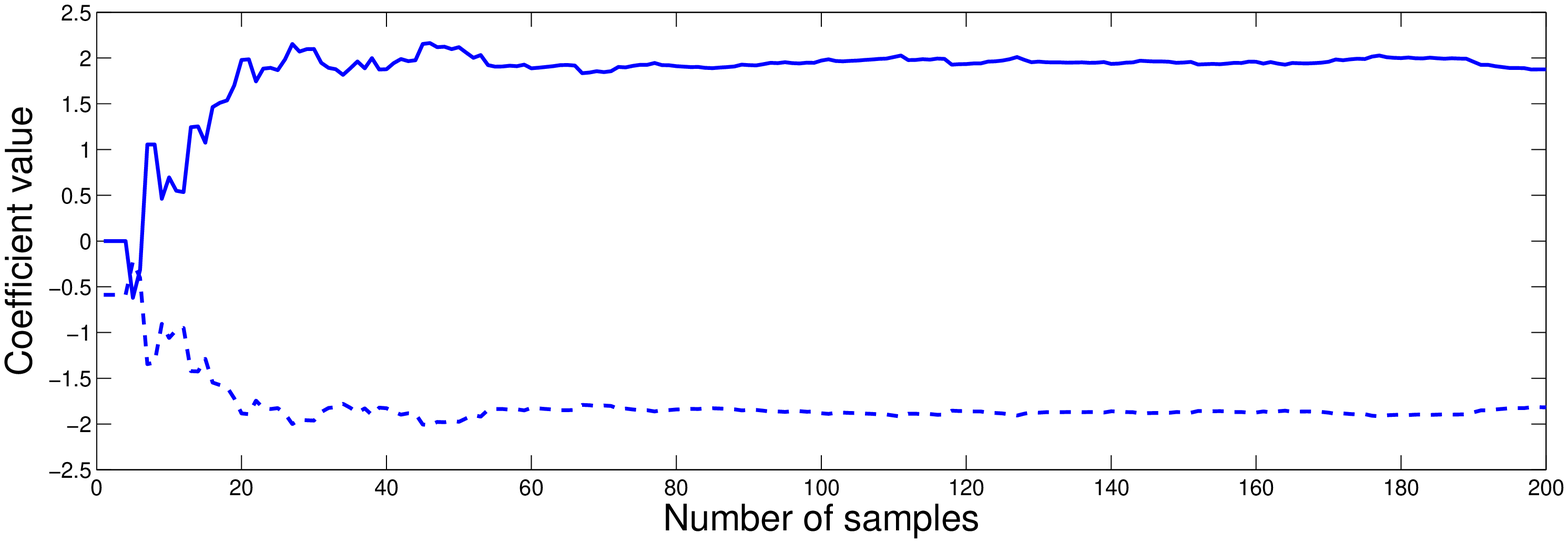}
         \caption{}
    \end{subfigure}   
    \begin{subfigure}[b]{0.5\textwidth}
        \includegraphics[width=8cm, height=3cm]{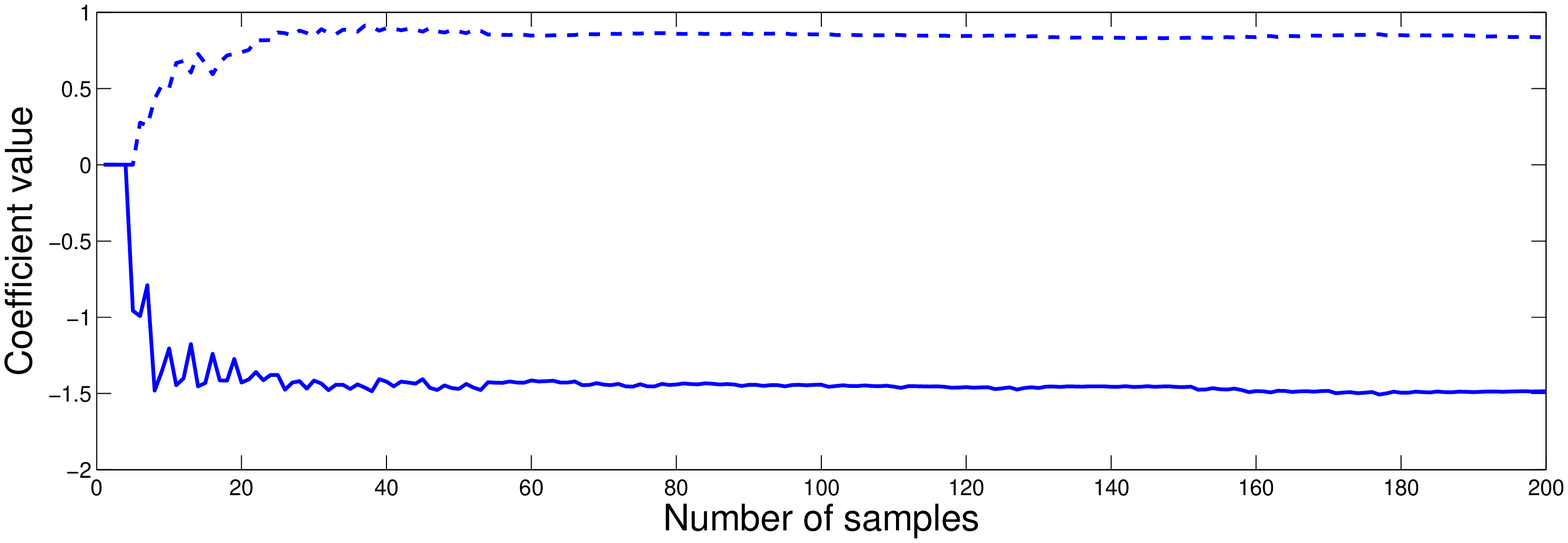}
        \caption{}
    \end{subfigure}
    \caption{For input signal of Experiment 2(iii), Parameter trajectory of filter coefficients (a)$\rm H_{0}$ and (b)$\rm H_{1}$.}
    \label{fig:convergence_propoerty(iii)}
\end{figure}

%

From the above results, the fast convergence property of the algorithm can be appreciated, for various kinds of input signals.

\subsection{}
We have  
observed that the proposed algorithm can be used to whiten Gaussian/Non-Gaussian processes with minimum as well as non-minimum phase. This observation is illustrated by considering the input and output spectra of the proposed SMWFB for different type of signals.

Experiment 5:
 We consider stochastic signals, with distributions given in column 1 of table \ref{table:input_signals}, filtered using three filters with transfer functions given in the first row of table \ref{table:input_signals}, the outputs, so obtained, are now used to design a two channel SMWFB. The spectra of input signal and output of the first channel of SMWFB are plotted in fig.(\ref{fig:convergence_propoerty(i)})-fig.(\ref{fig:convergence_propoerty(iii)}).
\begin{figure*}
\vspace{2cm}
\setcounter{temptablecounter1}{\value{figure}}
\begin{table}[H]
\begin{center}\begin{small}

\caption{Input Signals For Experiment 5}\label{table:input_signals}
\begin{tabular}{llll}
\hline 
 \multicolumn{1}{c}{Input signal}  & \multicolumn{1}{c}{Minimum phase} & \multicolumn{1}{c}{Maximum phase} &  \multicolumn{1}{c}{Mixed phase} \tabularnewline

Distribution & $\begin{aligned}H_{1}(z)= 1-
 0.8461z^{-1}\\ +0.9506z^{-2}
 \end{aligned}$ &$H_2(z)=\dfrac{z-1.2}{z^2-0.975z+0.9506}$& $H_3(z)=\dfrac{z^2-2.95z+1.90}{z^3-1.7750z^2+1.7306z-0.7605}$ \tabularnewline
 \hline
 Gaussian distributed,\\ $\mu=0$ 
 and ${\sigma}^2=1$  & Signal 1 & Signal 4 & Signal 7 \tabularnewline\hline
Uniformly distributed, \\ $$  [-1 1]  & Signal 2 & Signal 5   & Signal 8  \tabularnewline \hline
 Exponentially distributed ,\\  $\mu=1.5$ & Signal 3  & Signal 6   & Signal 9   \tabularnewline
\hline 
\end{tabular}
\end{small}\end{center}
\end{table}
\setcounter{figure}{\value{temptablecounter1}}
\end{figure*}
 
%

\begin{figure}[H]
 \begin{subfigure}[b]{0.3\textwidth}
    
       \includegraphics[width=8cm, height=2.5cm]{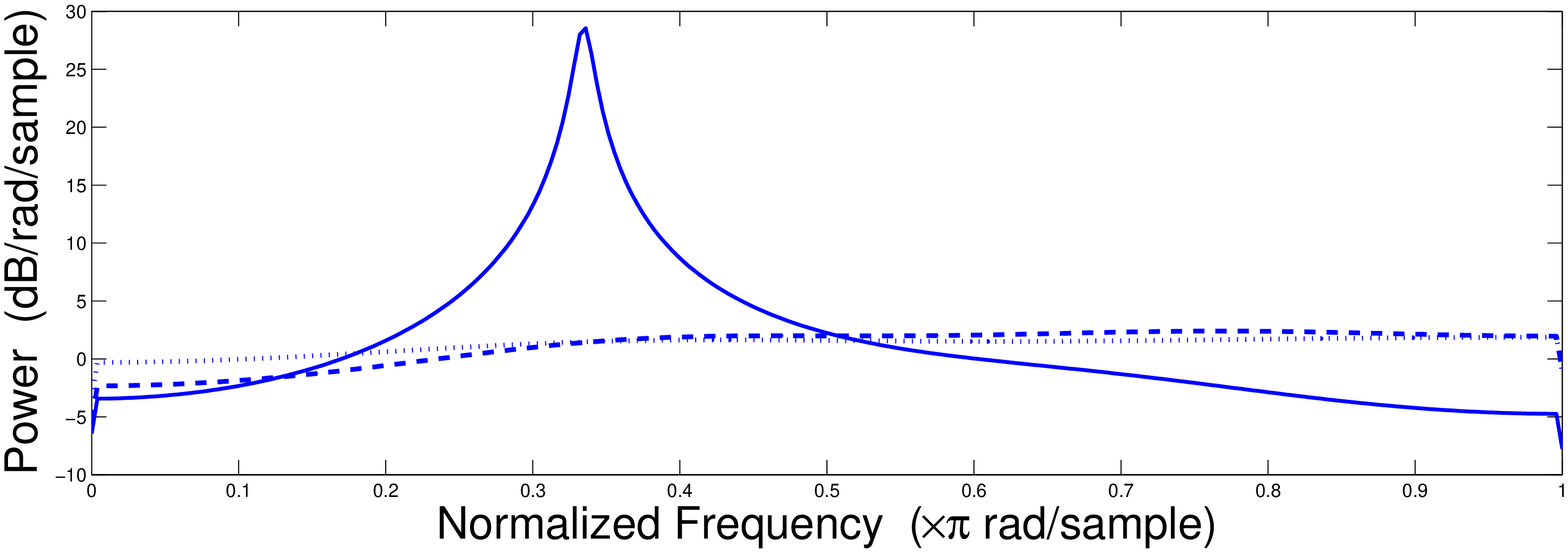}
        \caption{}
    \end{subfigure}
    \begin{subfigure}[b]{0.3\textwidth}
         \includegraphics[width=8cm, height=2.5cm]{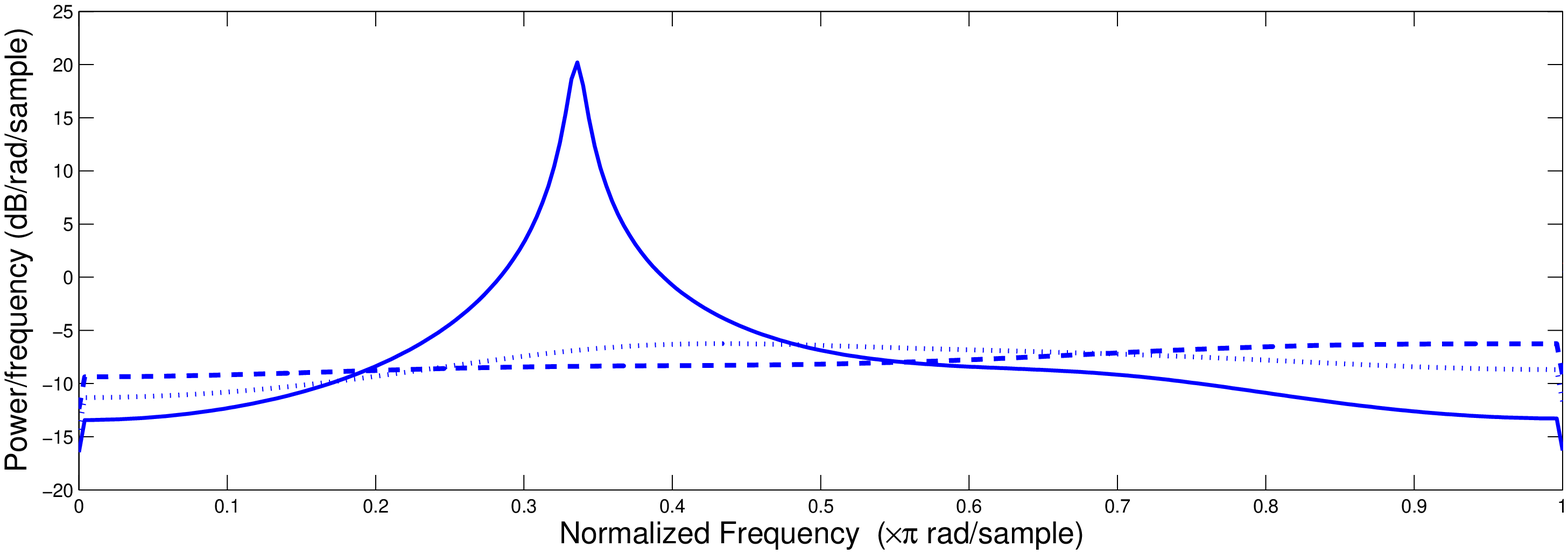}
        \caption{}
    \end{subfigure}   
    \caption{Power spectral density of input signal(solid line), Output of first channel(dashed line)
     and output of second channel(dotted line)
      when the input signal, as given in table \ref{table:input_signals} (a)signal 4, (b)signal 5.}
    \label{fig:whitening_max}
\end{figure}

\begin{figure}[H]
     \begin{subfigure}[t]{0.3\textwidth}
        \includegraphics[width=8cm, height=2.5cm]{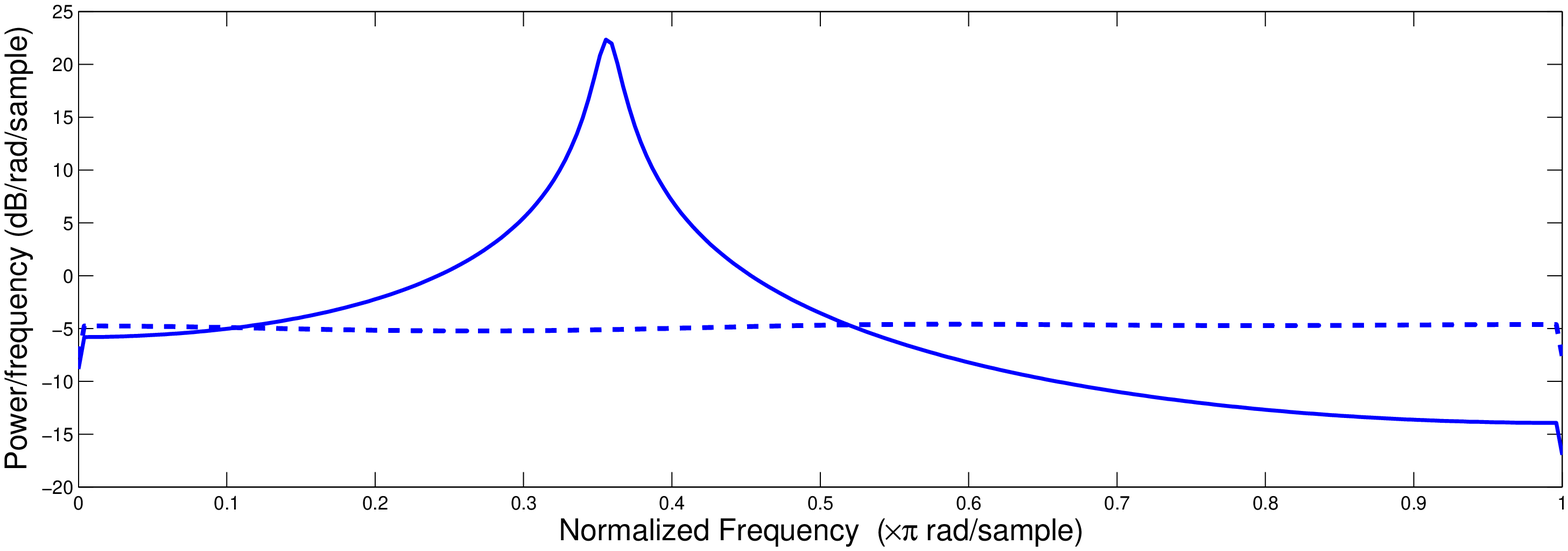}
        \caption{}
    \end{subfigure}
    \begin{subfigure}[b]{0.3\textwidth}
        \includegraphics[width=8cm, height=2.5cm]{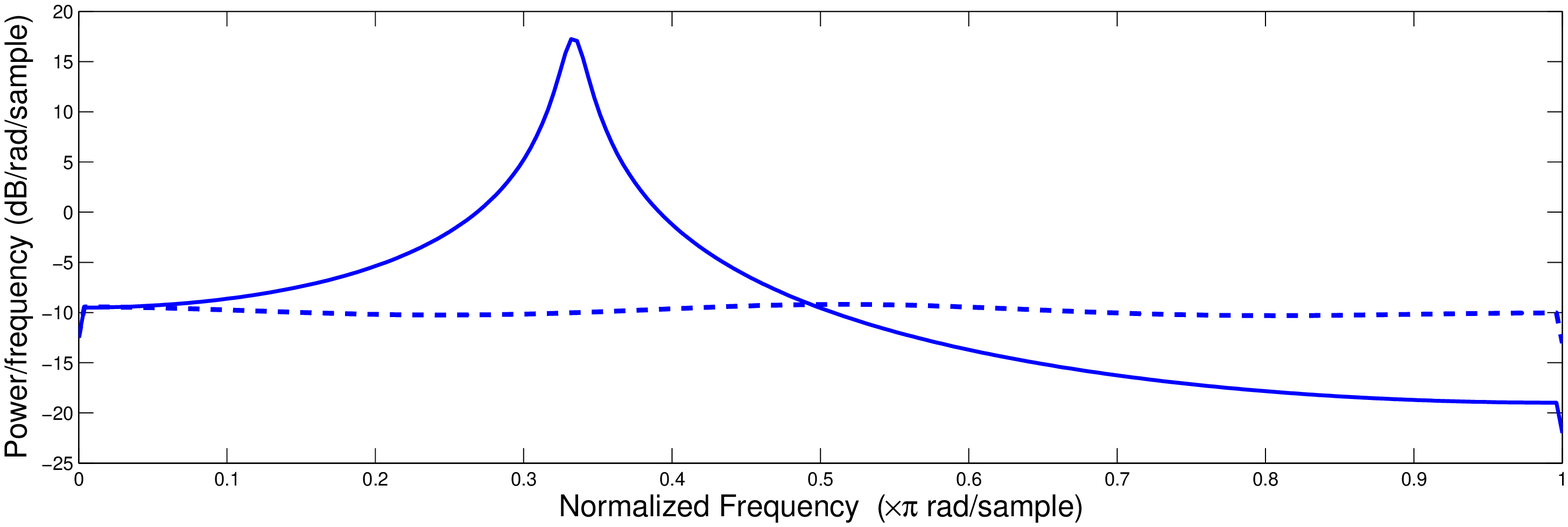}
        \caption{}
    \end{subfigure}   
    \begin{subfigure}[b]{0.3\textwidth}
                 
              \includegraphics[width=8cm, height=2.5cm]{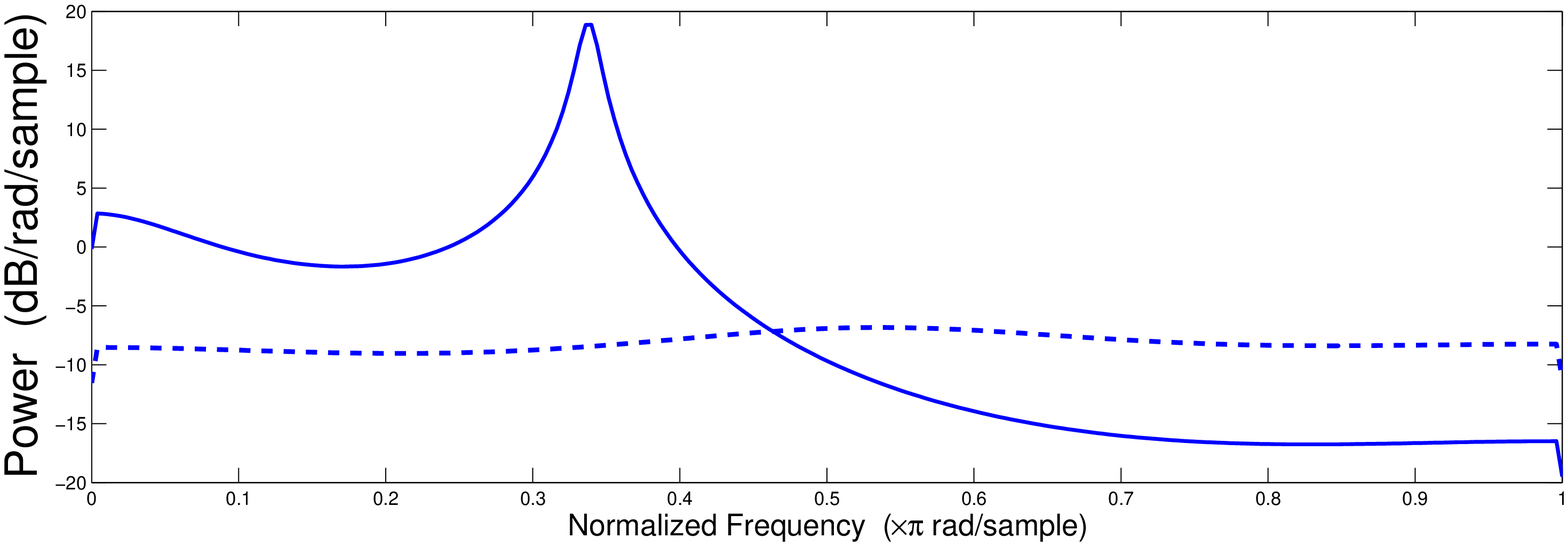}
        \caption{}
    \end{subfigure}
    \caption{Power spectral density of input signal(solid line),  Output of first channel(dashed line)
     and output of second channel(dotted line)
     when the input signal, as given in table \ref{table:input_signals} (a)signal 1, (b)signal 2 and (c)signal 3.}
    \label{fig:whitening_min}
\end{figure}


\begin{figure}[H]
 \begin{subfigure}[b]{0.5\textwidth}
     \includegraphics[width=8cm, height=2.5cm]{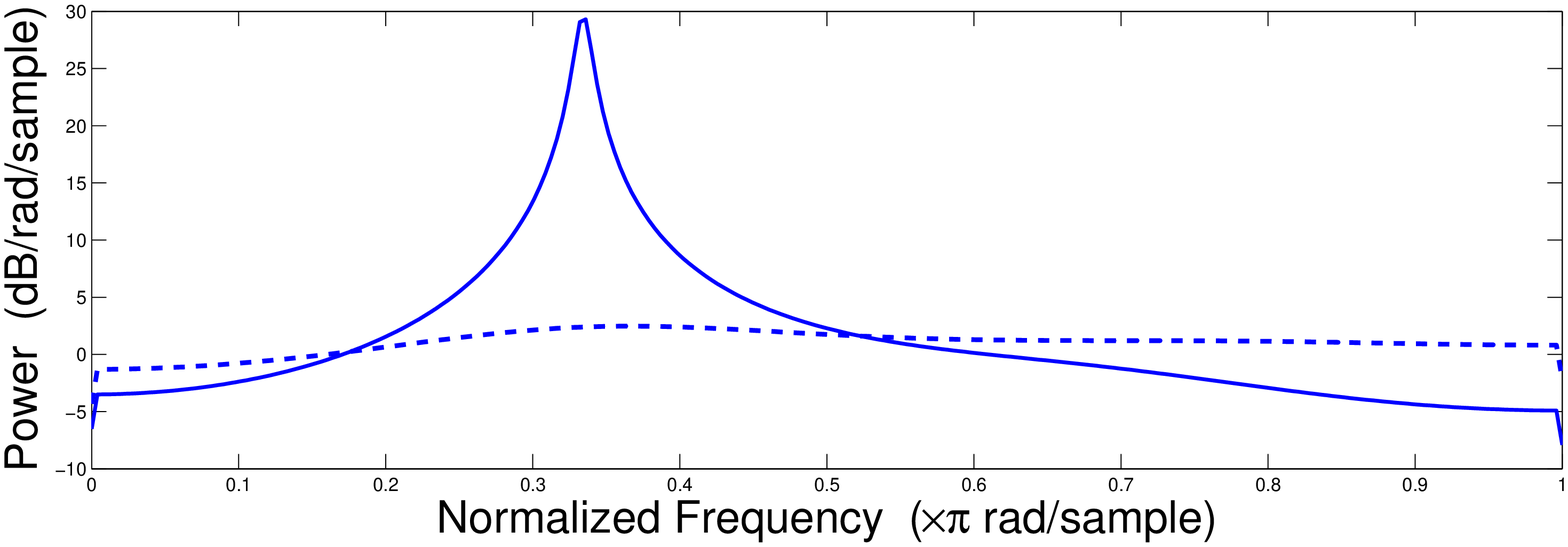}
        \caption{ }
    \end{subfigure}
    \begin{subfigure}[b]{0.5\textwidth}
       \includegraphics[width=8cm, height=2.5cm]{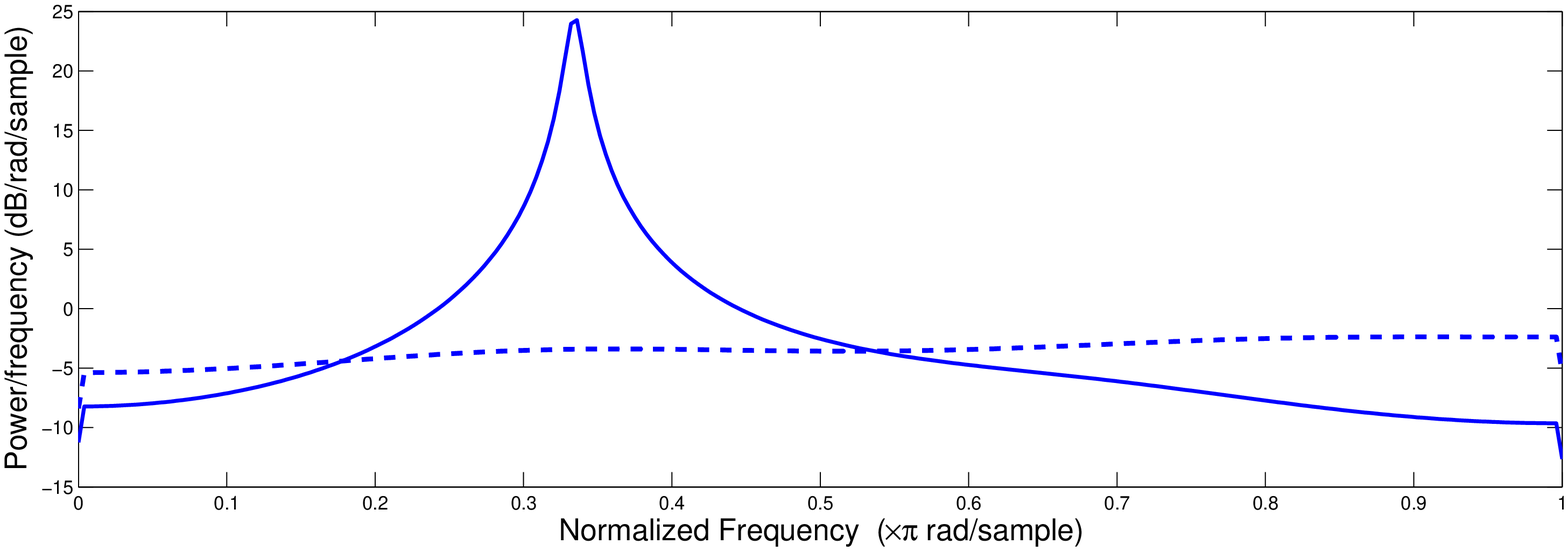}
        \caption{}
    \end{subfigure}   
    \begin{subfigure}[b]{0.3\textwidth}

       \includegraphics[width=8cm, height=2cm]{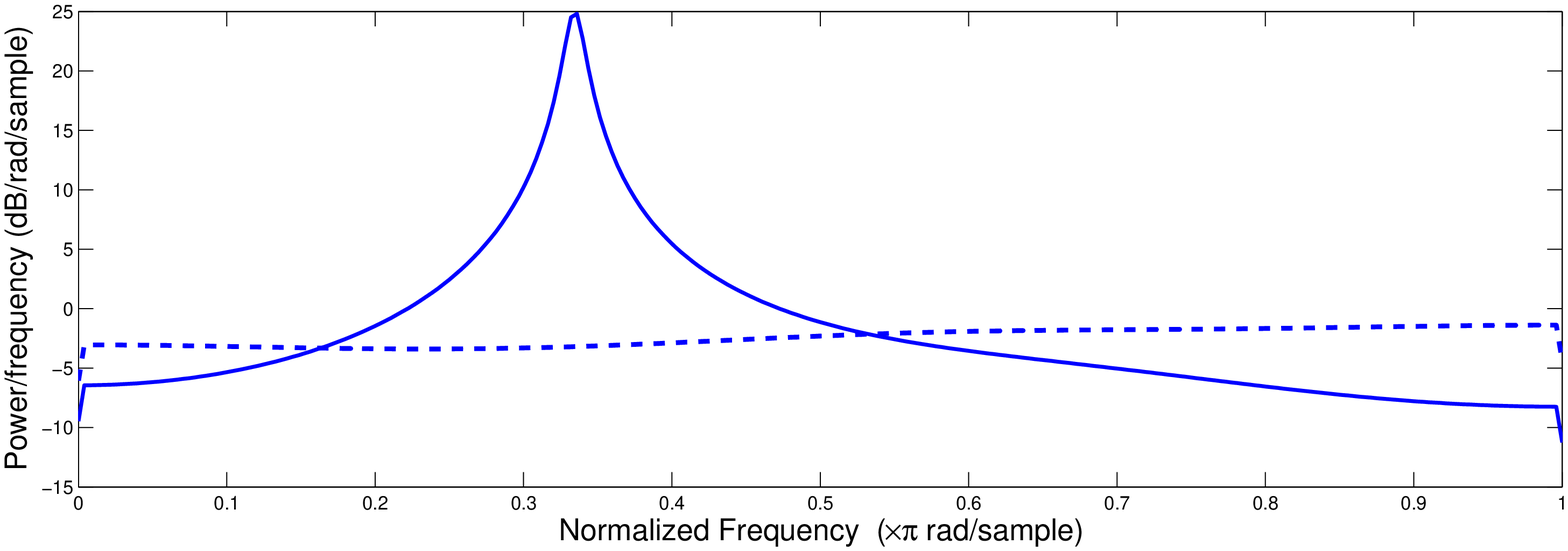}
        \caption{ }
    \end{subfigure}
    \caption{Power spectral density of input signal(solid line),  Output of first channel(dashed line)
     and output of second channel(dotted line)
      when the input signal, as given in table \ref{table:input_signals} (a)signal 7, (b)signal 8 and (c)signal 9.}
    \label{fig:whitening_mixed}
\end{figure}

\section{Conclusions}

In this paper we have proposed a fast least squares algorithm for a
modified signal matched multirate filter bank, with the aim of optimizing the coding
gain. It is also observed that the recursions of the algprithm gives rise toa lattice-ladder like structure. In table \ref{Table:cg_compared_Lu} and table \ref{table:cg_different_M}, we compare coding gain obtained using the proposed algorithm with some results present in the literature and a considerable improvement is obtained. Since we have not come across any other work dealing with adaptive algorithm in the context of signal adapted multirate filter banks, we compare our work with most recent block processing algorithms in the context. To obtain the optimized coding gain each channel is whitened across time and across bands, this is shown using simulation results, from fig.(\ref{fig:whitening_max}) , fig.(\ref{fig:whitening_min}) and fig.(\ref{fig:whitening_mixed}). From fig.(\ref{fig:whitening_max}), it can be noticed that the outputs of both channels have almost the same variance, a condition mentioned in section VI (A) for optimized coding gain.
Also, a least squares algorithm is presented for calculating order update recursions 
for the filter bank parameters, these parameters converge in 20-40 samples, to illustrate it we have plotted the convergence for three examples in
 fig.(\ref{fig:convergence_propoerty(i)}), fig(\ref{fig:convergence_propoerty(ii)}) and fig.(\ref{fig:convergence_propoerty(iii)}). The simulation results presented are for both Gaussian as well as non-Gaussian signals with minimum and non-minimum phase, as given in table \ref{table:input_signals}.
\section{Appendix}

In subsection A to D, some of the important relations used in this work, are presented to give a quick reference. In subsection E, we propose a lemma and a corollary to prove that the proposed algorithm in this paper orthogonalize the outputs across the channels.

\subsection{Least squares error in terms of projection operator}
Consider any row vector $\mathbf{\nu}$ and 
a matrix \begin{small} $\rm {V_{1:n}}=\left[\begin{array}{cccc}
\mathbf{\nu_{1}}^T & 
\mathbf{\nu_{2}}^T &
\cdots &
\mathbf{\nu_{n}}^T
\end{array}\right]^{T}$\end{small} , where $
\mathbf{\nu_{i}}$'s are row vectors $\in$ $\mathbb{R}^{L}$, L $\geq n$. Consider $\mathbf{e}$ as the error in estimating $\mathbf{\nu}$ from the rows of $\rm {V_{1:n}}$, given as:

\begin{align}
\mathbf{e}=\mathbf{\nu} + aV_{1:n} \label{eq:LSerror}
\end{align}

where, $'a'$ is the parameter vector. If $\mathbf{e}$ is minimized in the least squares sense, 
implies $\mathbf{e}$ is orthogonal to space span $\{\rm {V_{1:n}}\}$, the optimum '$a$' can be written as:
\begin{align}
a= -\nu V^{T}{(V_{1:n}V_{1:n}^{T})}^{-1}\label{eq:least_squres_par_est}
\end{align}

Substituting the value of '$a$' in (\ref{eq:LSerror}), we get:
\begin{align*}
e=\nu-\nu V^{T}{(V_{1:n}V_{1:n}^{T})}^{-1}V_{1:n}
\end{align*}
Denoting $V^{T}(VV^{T})^{-1}V$ in terms of projection operator as $P\left[V\right]$, we can the re-write the above equation as:

\begin{align}
e=\nu -\nu P\left[V_{1:n}\right]\nonumber\\
e=\nu P^{\perp}\left[V_{1:n}\right]\label{eq:Projection}
\end{align}
 where, $P^{\perp}$ denotes the projection on the orthogonal complement space of V. From the above equation, it cab be clearly observed that the least squares error lies in the orthogonal complement space spanned by the rows of span $\rm {V_{1:n}}$. 

\subsection{Inner Product Update Formula:} 

For any two row vectors $\mathbf{\nu}$ and $\mathbf{w}$ and the above defined $\rm{V_{1:n}}$ the inner product update relation as given in \cite{Friedlander} is:

\begin{eqnarray}
 && \hspace*{-1.5cm}\mathbf{\nu}\rm P^{\perp}\left[V_{1:n} \right]\mathbf{w}^T =
 \mathbf{\nu}\rm P^{\perp}\left[V_{1:n-1} \right]\mathbf{w}^T- 
 \mathbf{\nu} \rm  P^{\perp}\left[V_{1:n-1} \right]\mathbf{\nu}_n^T \nonumber\\ &&\hspace*{-1.5cm}
 \left[\mathbf{\nu}_n \rm P^{\perp}[V_{1:n-1}\mathbf{\nu}_n^T]^{-1}\right].
\mathbf{\nu}_n \rm P^{\perp} \left[V_{1:n-1} \right] \mathbf{w}^T \label{eq:inner_product_formula}
\end{eqnarray}

\subsection{Corollary 1.3 from \cite{joshiI}}
Let $\mathbf{\nu}$, $\mathbf{w}$ and $\mathbf{x}$ be row vectors belonging to the space $\mathbb{R^{L}}$ and $\rm {V_{1:n}}$ is the matrix defined above. In \cite{joshiII}, corollary 1.3 is given as:

\begin{eqnarray}
&& \hspace*{-1.5cm}\mathbf{\nu}\rm P^{\perp}\left[\mathbf{x}\right] \rm P^{\perp}\left[V_{1:n}P^{\perp}\left[\mathbf{x}\right]\right]\mathbf{w}^T \nonumber\\&&
=\mathbf{\nu} \rm P^{\perp}\left[V_{1:n}\right]\mathbf{w}^T- \mathbf{\nu} \rm P\left[\mathbf{x} \rm P^{\perp}\left[V_{1:n}\right]\right]\mathbf{w}^T\label{eq:corollary}
\end{eqnarray}

\subsection{Pseudo-Inverse Update Relation from \cite{joshiII}:}

Let $\mathbf{z}$ and $\mathbf{x}$ be row vectors $\in \mathbb{R^{M}}$ and define\\
 $\begin{small} \rm V_{1:n}[\mathbf{x}]= \left[\rm{(P^{\perp}[\mathbf{x}]\mathbf{\nu}_1^T)} \cdots  \rm{(P^{\perp}[\mathbf{x}]\mathbf{\nu}_n^T)} \right]^{T}\end{small}$ and\\
    $\rm {K_{n}[\mathbf{x}]=\rm V_{1:n}^{T}[\mathbf{x}][\rm V_{1:n}[\mathbf{x}]\rm V_{1:n}^{T}[\mathbf{x}]]^{-1}}$, 
 which is the pseudo-inverse of $\rm {V_{1:n}[\mathbf{x}]}$.
 \cite{joshiII} gives an update relation for pseudo-inverse i.e., $\rm{K_{n}[\mathbf{x}]}$ as:

\begin{eqnarray}
{\mathbf{z} \rm K_n[\mathbf{x}]}=[{\mathbf{z} \rm K_{n-1}[\mathbf{0}]}\mid 0]
 +{\mathbf{z} \rm P^{\perp}[V_{1:n-1}] \mathbf{\nu}_n^T} \nonumber\\
 {[\mathbf{\nu}_n \rm P^{\perp}[V_{1:n-1}] \mathbf{\nu}_n^T]^{-1}}
.{[(-\mathbf{\nu}_{n} \rm K_{n-1}[\mathbf{0}])}\mid 1] \nonumber\\
+{\mathbf{z} \rm P^{\perp}[V_{1:n}] \mathbf{x}^T[ \mathbf{x} \rm P^{\perp}[V_{1:n}] \mathbf{x}^T]^{-1}}.
 {(-\mathbf{x} \rm K_n[0])}\label{eq:pseudo_inverse_update}
\end{eqnarray}
 It can be observed from the above update equation that the first two terms of the right hand side will update the matrix ${K_{n-1}[0]}$ to ${K_{n}[0]}$ and the last term updates ${K_{n}[0]}$ to ${K_{n}[x]}$.

\subsection{Lemma 2 and Corollary 1}

\begin{lem}
\begin{equation}
\rm P\left[{V_{1:n}}P^{\perp} \left[ \mathbf{x} \rm  P^{\perp}\left[{V_{1:n}}\right]\right]\right]=\rm P\left[{V_{1:n}}\right]\label{eq:20}
\end{equation}
where, 
 $\mathbf{x}$ are any row vector $\in$ $\mathbb{R}^{L}$.
\end{lem}

Proof:
In \cite{joshiI} an augmented set has been defined similarly an augmented set $\left[\rm {V_{1:n}}\mid \mathbf{x} \rm P^{\perp}\left[{V_{1:n}}\right]\right]$ is defined here 
as:
\begin{small}\begin{equation}
\left[\rm {V_{1:n}}\mid \mathbf{x} \rm P^{\perp}\left[{V_{1:n}}\right]\right]=\left[\begin{array}{c}
\mathbf{\nu_{1}}\\
\mathbf{\nu_{2}}\\
\vdots\\
\mathbf{\nu_{n}}\\
\mathbf{x}\rm P^{\perp}\left[{V_{1:n}}\right]
\end{array}\right].\label{eq:21}
\end{equation}\end{small}

Let $S\left[\rm {V_{1:n}}\mid \mathbf{x} \rm P^{\perp}\left[{V_{1:n}}\right]\right]$ denote
the space spanned by the row vectors of $\left[\rm {V_{1:n}}\mid \mathbf{x} \rm P^{\perp}\left[{V_{1:n}}\right]\right]$.
Therefore, this space can be written as:
\begin{eqnarray}
S\left[\rm {V_{1:n}}\mid \mathbf{x} \rm P^{\perp}\left[{V_{1:n}}\right]\right]=S\left[{V_{1:n}} \rm P^{\perp}\left[\mathbf{x}\rm P^{\perp}\left[{V_{1:n}}\right]\right]\right] \nonumber\\
\oplus S\left[\mathbf{x} \rm P^{\perp}\left[{V_{1:n}}\right]\right]\label{eq:22}
\end{eqnarray}

and also as
 \begin{eqnarray}
S\left[\rm {V_{1:n}}\mid \mathbf{x} \rm P^{\perp}\left[{V_{1:n}}\right]\right]=S\left[{V_{1:n}}\right]\oplus S\left[\mathbf{x}\rm P^{\perp}\left[{V_{1:n}}\right]\right]\label{eq:23}
\end{eqnarray}

Comparing (\ref{eq:22}) and (\ref{eq:23}) we get:
\begin{equation}
S\left[\rm V_{1:n}P^{\perp} \mathbf{x} \rm P^{\perp}\left[{V_{1:n}}\right]\right]=S\left[{V_{1:n}}\right]\label{eq:24}
\end{equation}

The above equation can be written in terms if projection operator
as:
\begin{equation}
\rm P \left[{V_{1:n}}P^{\perp} \mathbf{x} \rm P^{\perp}\left[{V_{1:n}}\right]\right]=\rm P\left[{V_{1:n}}\right]\label{eq:25}
\end{equation}
 \hspace{8.5cm} $\blacksquare$

\newtheorem{cor}{Corollary}
\begin{cor}

\begin{eqnarray}
\mathbf{\nu}\rm P^{\perp}\left[\mathbf{x}\rm P^{\perp}
\left[\rm {V_{1:n}}\right]\right]
\rm P^{\perp}\left[\rm {V_{1:n}}\rm P^{\perp}\left[\mathbf{x}\rm P^{\perp}\left[\rm {V_{1:n}}\right]\right]\right]\mathbf{w}^{T}\nonumber\\=
\mathbf{\nu} \rm P^{\perp}\left[\mathbf{x}\right]\rm P^{\perp}\left[\rm {V_{1:n}}\rm P^{\perp}\left[\mathbf{x}\right]\right]\mathbf{w}^{T}\label{eq:col1}
\end{eqnarray}
 
 \end{cor}

Proof:
LHS of (\ref{eq:col1}) is:

\begin{small}
\begin{eqnarray}
& &\hspace{-0.65cm} \mathbf{\nu} \rm P ^{\perp}\left[\mathbf{x}\rm P^{\perp}\left[{V_{1:n}}\right]\right]\rm P^{\perp}\left[{V_{1:n}}\rm P^{\perp}\left[\mathbf{x}\rm P^{\perp}\left[{V_{1:n}}\right]\right]\right]\mathbf{w}^{T}\nonumber \\
&&\hspace{-0.65cm} =\mathbf{\nu}\left({I}-\rm P\left[\mathbf{x} \rm P ^{\perp}\left[{V_{1:n}}\right]\right]\right)\left({I}-\rm P\left[{V_{1:n}}\rm P^{\perp}\left[\mathbf{x}\rm P^{\perp}\left[{V_{1:n}}\right]\right]\right]\right)\mathbf{w}^{T} \nonumber\\
&&\hspace{-0.65cm} =\mathbf{\nu} \rm P^{\perp}\left[{V_{1:n}} \rm P^{\perp}\left[\mathbf{x} \rm P^{\perp}\left[{V_{1:n}}\right]\right]\right]\mathbf{w}^{T}-\mathbf{\nu}\rm P\left[\mathbf{x}\rm P^{\perp}\left[{V_{1:n}}\right]\right]\mathbf{w}^{T}\label{eq:26}
\end{eqnarray}\end{small}

as, $\mathbf{\nu}\rm P\left[\mathbf{x}\rm P^{\perp}\left[{V_{1:n}}\right]\right]\rm P\left[{V_{1:n}} \rm P^{\perp}\left[\mathbf{x}\rm P^{\perp}\left[{V_{1:n}}\right]\right]\right]$ is a zero operator.

Form (\ref{eq:26}) and ({\ref{eq:25}) :
\begin{eqnarray}
\mathbf{\nu} \rm P^{\perp}\left[\mathbf{x} \rm P^{\perp}\left[{V_{1:n}}\right]\right]\rm P^{\perp}\left[{V_{1:n}} \rm P^{\perp}\left[{V_{1:n}}\right]\right]\mathbf{w}^{T}\nonumber\\
=\mathbf{\nu} \rm P^{\perp}\left[{V_{1:n}}\right]\mathbf{w}^{T}-\mathbf{\nu} \rm P\left[\mathbf{x}\rm P^{\perp}\left[{V_{1:n}}\right]\right]\mathbf{w}^{T}
\\
=\mathbf{\nu} \rm P^{\perp}\left[\mathbf{x}\right]\rm P^{\perp}\left[{V_{1:n}}\rm P^{\perp}\left[\mathbf{x}\right]\right]\mathbf{w^{T}}\nonumber
\end{eqnarray}
 \hspace{8.5cm} $\blacksquare$

In order to understand how the filter A will orthogonalize the channels, substitute the following values in the LHS of (\ref{eq:21}):\\
$\mathbf{\nu}=\mathbf{x}(Mn-i)$, $\rm V_{1:n}=X_{p}^{Mn-M}$, $\mathbf{x}=\rm {{X}^{Mn-i-1}_{M-1-i}}$ and $\mathbf{w}=\mathbf{\pi}$ 
we get:

\begin{small}
\begin{align}
&  \mathbf{x}(Mn-i)\rm {P^{\perp}\left[{X}^{Mn-i-1}_{M-1-i}P^{\perp}\left[{X}^{Mn-M}_{p}\right]\right]} \nonumber\\
&\rm{P^{\perp}\left[{X}^{Mn-M}_{p}P^{\perp}\left[{X}^{Mn-i-1}_{M-1-i}P^{\perp}\left[{X}^{Mn-M}_{p}\right]\right]\right]\pi^T}\nonumber\\
&  =\mathbf{x}(Mn-i)\rm{P^{\perp}\left[{X}^{Mn-i-1}_{M-1-i}\right]}
 \rm{P^{\perp}\left[{X}^{Mn-M}_{p}P^{\perp}\left[{X}^{Mn-i-1}_{M-1-i}\right]\right]\pi^T}\nonumber
\end{align}
\end{small}

\bibliographystyle{IEEEtran}
\bibliography{IEEEabrv,references}

\begin{thebibliography}{10}
\providecommand{\url}[1]{#1}
\csname url@samestyle\endcsname
\providecommand{\newblock}{\relax}
\providecommand{\bibinfo}[2]{#2}
\providecommand{\BIBentrySTDinterwordspacing}{\spaceskip=0pt\relax}
\providecommand{\BIBentryALTinterwordstretchfactor}{4}
\providecommand{\BIBentryALTinterwordspacing}{\spaceskip=\fontdimen2\font plus
\BIBentryALTinterwordstretchfactor\fontdimen3\font minus
  \fontdimen4\font\relax}
\providecommand{\BIBforeignlanguage}[2]{{%
\expandafter\ifx\csname l@#1\endcsname\relax
\typeout{** WARNING: IEEEtran.bst: No hyphenation pattern has been}%
\typeout{** loaded for the language `#1'. Using the pattern for}%
\typeout{** the default language instead.}%
\else
\language=\csname l@#1\endcsname
\fi
#2}}
\providecommand{\BIBdecl}{\relax}
\BIBdecl

\bibitem{Friedlander}
B.~Friedlander, ``Lattice filters for adaptive processing,'' \emph{Proc. of
  IEEE}, vol.~70, no.~8, pp. 829--867, 1982.

\bibitem{Nalbalwar_phd}
S.~Nalbalwar, ``Some studies on signal matched multirate filter bank,'' Ph.D.
  dissertation, IITD, 2008.

\bibitem{ref1}
M.~K. Tsatsanis and G.~B. Giannakis, ``Principal component filter banks for
  optimal multiresolution analysis,'' \emph{IEEE Trans. Signal Processing},
  vol.~43, no.~8, pp. 1766--1777, 1995.

\bibitem{Lu}
W.~S. Lu and A.~Antoniou, ``Design of signal-adapted biorthogonal filter
  banks,'' \emph{IEEE Trans. Circuits and Systems I: Fundamental Theory and
  Applications}, vol.~48, no.~1, pp. 90--102, 2001.

\bibitem{weng}
C.~C. Weng and P.~P. Vaidyanathan, ``The role of gtd in optimizing perfect
  reconstruction filter banks,'' \emph{IEEE Trans. Signal Processing}, vol.~60,
  no.~1, pp. 112--128, 2012.

\bibitem{A.gupta}
A.~Gupta, S.~D. Joshi, and S.~Prasad, ``A new approach for estimation of
  statistically matched wavelet,'' \emph{IEEE Trans. Signal Processing},
  vol.~53, no.~5, pp. 1778--1793, 2005.

\bibitem{Delsarte}
P.~Desarte, B.~Macq, and D.~T.~M. Slock, ``Signal-adapted multiresolution
  transform for image coding,'' \emph{IEEE Trans. Information Theory}, vol.~38,
  no.~2, pp. 897--904, 1992.

\bibitem{ref3}
J.~W. Woods and S.~D. O'Neil, ``Subband coding of images,'' \emph{IEEE Trans.
  Acoustics, Speech and Signal Processing}, vol.~34, no.~5, pp. 1278--1288,
  1986.

\bibitem{ref4}
R.~Cox, D.~Bock, K.~Bauer, J.~D. Johnston, and J.~H. Snyder, ``The analog voice
  privacy system,'' in \emph{Acoustics, Speech, and Signal Processing, IEEE
  International Conference on ICASSP '86.}, vol.~11, 1986, pp. 341--344.

\bibitem{gadre}
A.~Jhawar, P.~Ginde, P.~Patwardhan, and V.~M. Gadre, ``Coding gain optimized
  finite impulse response (fir) paraunitary (pu) filter banks,'' in
  \emph{National Conference on Communications (NCC), 2010}, 2010, pp. 1--5.

\bibitem{joshiI}
S.~Prasad and S.~D. Joshi, ``A new recursive pseudo least squares algorithm for
  {ARMA} filtering and modeling. i,'' \emph{IEEE Trans. Signal Processing},
  vol.~40, no.~11, pp. 2766--2774, 1992.

\bibitem{joshiII}
------, ``A new recursive pseudo least squares algorithm for arma filtering and
  modeling. ii,'' \emph{IEEE Trans. Signal Processing}, vol.~40, no.~11, pp.
  2775--2783, 1992.

\bibitem{Vaidyanathan}
P.~P. Vaidyanathan, \emph{Multirate systems and filter banks}.\hskip 1em plus
  0.5em minus 0.4em\relax Upper Saddle River, NJ, USA: Prentice-Hall, Inc.,
  1993.

\bibitem{Katto}
J.~Katto and Y.~Yasuda, ``Performance evaluation of subband coding and
  optimization of its filter coefficients,'' \emph{Jornal of Visual
  Communication and Image Representation}, vol.~2, no.~4, pp. 303--313, 1991.

\bibitem{Aase}
S.~Aase and T.~Ramstad, ``On the optimality of nonunitary filter banks in
  subband coders,'' \emph{Image Processing, IEEE Transactions on}, vol.~4,
  no.~12, pp. 1585--1591, Dec 1995.

\bibitem{PLT}
S.-M. Phoong and Y.-P. Lin, ``Prediction-based lower triangular transform,''
  \emph{Signal Processing, IEEE Transactions on}, vol.~48, no.~7, pp.
  1947--1955, Jul 2000.

\end{thebibliography}

 \end{document}